\documentclass{aa}  

\usepackage{xcolor}
\usepackage{graphicx}
\usepackage{txfonts}
\usepackage{textcomp}
\usepackage{hyperref}
\newcommand{\angstrom}{\text{\normalfont\AA}}
\begin{document}

   \title{Polarization aberrations in next-generation giant segmented mirror telescopes (GSMTs)}

   \subtitle{I.Effect on the coronagraphic performance}
   
   \author{Ramya M Anche\inst{1}
   \and Jaren N Ashcraft\inst{1,2}
    \and Sebastiaan Y. Haffert \inst{1}\fnmsep\thanks{NASA Hubble Fellow}
    \and Maxwell A. Millar-Blanchaer \inst{3}
    \and Ewan S. Douglas\inst{1}
    \and Frans Snik\inst{4}
    \and Grant Williams\inst{1}
    \and Rob G. van Holstein\inst{5}
    \and David Doelman\inst{4}
    \and Kyle Van Gorkom\inst{1}
    \and Warren Skidmore\inst{6}}

   \institute{Steward Observatory, University of Arizona, 933N Cherry Avenue, Tucson, Arizona, 85721, USA \\ 
    \email{ramyaanche@arizona.edu}
         \and James C. Wyant College of Optical Sciences, University of Arizona, 933N Cherry Avenue, Tucson, Arizona, 85721, USA        
         \and Department of Physics, University of California, Santa Barbara, CA, 93106, USA
         \and Leiden University, Niels Bohrweg 2, 2333 CA Leiden 
         \and European Southern Observatory (ESO), Alonso de Córdova 3107, Vitacura, Casilla 19001, Santiago de Chile, Chile
         \and TMT International Observatory LLC, 100 W. Walnut St., Suite 300, Pasadena, CA 91124, USA  }

   \date{Received 7 December 2022 / Accepted 5 February 2023}

  \abstract  
  {Next-generation large segmented mirror telescopes are expected to perform direct imaging and characterization of Earth-like rocky planets, which requires contrast limits of $10^{-7}$ to $10^{-8}$ at wavelengths from I to J band. One critical aspect affecting the raw on-sky contrast are polarization aberrations (i.e., polarization-dependent phase and amplitude patterns in the pupil) arising from the reflection from the telescope's mirror surfaces and instrument optics. These polarization aberrations induce false signals for polarimetry that can be calibrated to a certain degree, but they can also fundamentally limit the achievable contrast of coronagraphic systems.}
  {We simulate the polarization aberrations and estimate their effect on the achievable contrast for three next-generation ground-based large segmented mirror telescopes.}
   {We performed ray-tracing in Zemax\textsuperscript{\textregistered} and computed the polarization aberrations and Jones pupil maps using the polarization ray-tracing algorithm. The impact of these aberrations on the contrast is estimated by propagating the Jones pupil maps through a set of idealized coronagraphs using hcipy, a physical optics-based simulation framework.}
   {The optical modeling of the giant segmented mirror
telescopes (GSMTs) shows that polarization aberrations create significant leakage through a coronagraphic system. The dominant aberration is retardance defocus, which originates from the steep angles on the primary and secondary mirrors. The retardance defocus limits the contrast to $10^{-5}$ to $10^{-4}$ at 1 $\lambda/D$ at visible wavelengths, and $10^{-5}$ to $10^{-6}$ at infrared wavelengths. The simulations also show that the coating plays a major role in determining the strength of the aberrations.}
   {Polarization aberrations will need to be considered during the design of high-contrast imaging instruments for the next generation of extremely large telescopes. This can be achieved either through compensation optics, robust coronagraphs, specialized coatings, calibration, and data analysis approaches, or by incorporating polarimetry with high-contrast imaging to measure these effects.}


   \maketitle
%

\section{Introduction}
The polarization-dependent phase and amplitude pupil patterns originating from differences in the Fresnel coefficients (s and p polarization states) are called polarization aberrations \citep{mcguire1987polarization,chipman1987polarization}. The polarization aberration theory to estimate these polarization aberrations and their effect on the point spread function (PSF) of optical systems has been very well developed and presented over the last two decades \citep{chipman1989polarization,mcguire1988polarization,mcguire1990diffraction,mcguire1994polarization,sanchez1992instrumental,sanchez1994instrumental,breckinridge2015polarization}. 
\par
\cite{chipman2018polarized} describes the dominant polarization aberrations in optical systems to be diattenuation piston, tilt, and defocus, and retardance piston, tilt, and defocus. Diattenuation is a polarization-dependent amplitude apodization calculated using the difference in maximum and minimum reflection or transmission through an interface. Retardance is a polarization-dependent phase aberration calculated from the difference in the maximally and minimally phase-delayed polarization state. The change in diattenuation and retardance can be expressed in elementary vector\footnote{Technically, the axis of diattenuation and retardance repeats every $\pi$ rotation, rather than $2\pi$ like a typical vector. Therefore, \cite{chipman2018polarized} refers to them as \emph{directors}.} shapes that show the orientation and magnitude of these aberrations. The diattenuation- and retardance-piston axes are of uniform magnitude across the pupil and oriented in the same direction. The diattenuation- and retardance-tilt axes rotate by $\pi/2$ around the pupil, and the magnitude increases linearly. The axes of diattenuation- and retardance-defocus rotate by $\pi$ around the pupil, and the magnitude increases quadratically. We refer the readers to chapter 10 of \cite{chipman2018polarized} to the maps of these polarization aberrations. 
\par In the context of telescopes, for an unpolarized light incident on a Cassegrain telescope with a fold mirror, quadratic variation in phase and amplitude of the reflection coefficients from primary and secondary mirrors causes the polarization in $X$ and $Y$ to be astigmatic, giving rise to diattenuation- and retardance-defocus. In addition, the fold mirror's linear phase and amplitude variation cause diattenuation piston and tilt, and retardance piston and tilt, leading to a beam shift in the final image \citep{van2021high}.
The polarization-induced on-axis astigmatism and chromatic aberration were estimated for an F/1.5 Cassegrain telescope by \cite{reiley1992coating}. The polarization aberrations were evaluated for the Solar Activity Measurements Experiments (SAMEX) Solar Vector Magnetograph by \cite{mcguire1989polarization}, and significant improvement was obtained in the polarization accuracy by minimizing the angle of incidence and the difference between the reflection coefficients. \cite{clark2011polarization} shows the compensation of Fresnel aberrations arising in low F-number telescopes due to mirror curvature and coating.  
\par Recently, \cite{breckinridge2015polarization} and \cite{chipman2015polarization} estimated the polarization aberrations for a 2.4 m Cassegrain telescope with a fold mirror to understand the magnitude of these aberrations and their effect on the exoplanet measurements. The diattenuation aberrations cause amplitude apodization, but have a smaller effect on the image quality, while the retardance aberrations (piston, defocus, tilt) give rise to a ghost PSF, leading to the ellipticity of the Airy disk. The intensity of the ghost PSF is estimated to be about one part in $10^{-4}$ of the two primary PSF images for the $X$- and $Y$-polarized light. Though the magnitude is small, the ghost PSF has a complex structure and larger spatial extent. For an unpolarized star, the orthogonally polarized components ($X$ and $Y$) have different polarization aberrations, and a regular adaptive optics system cannot optimally correct these aberrations simultaneously. Thus, in a coronagraphic system designed for exoplanet studies, the ghost PSF may overfill the focal plane mask and result in the burying of the terrestrial exoplanet signal. 
\par
The observations from the high-contrast imaging instruments at the ground-based large telescopes have already shown the beam shift introduced as a consequence of the retardance tilt from the telescope mirrors. The polarization aberrations caused by the fold mirror and the derotator system in the Very Large Telescope (VLT) give rise to a differential polarimetric beam shift of about 1 mas (57.3nm) between the orthogonal polarization components, which manifests as positive and negative features on the opposite sides of the stellar PSF \citep{schmid2018sphere}. For the Gemini telescope, \cite {millar2022polarization} have observed a retardance defocus of $\sim$3 nm between incident X and Y polarization states in GPI polarization observations. It has become evident that, although the magnitude of the polarization aberrations is small, they cause significant errors in the coronagraphic performance and bias high-contrast polarimetric measurements \citep{safonov2022differential} and  \citep{millar2022polarization}. Thus, the polarization aberrations are now modeled for some of the telescopes with a high-contrast imaging instrument to understand their effect on the achievable contrast and for designing the mitigation or calibration strategy \citep{breckinridge2018terrestrial},\citep{krist2018wfirst}, \citep{mendillo2019polarization},\citep{will2019effects} and \citep{van2020calibration}. 
\par 
The next generation of giant segmented mirror telescopes (GSMTs),  \href{https://www.eso.org/sci/facilities/eelt/}{the Extremely Large Telescope (ELT)}, \href{https://www.tmt.org/}{the Thirty Meter Telescope (TMT)}, and  \href{https://www.gmto.org/}{the Giant Magellan Telescope (GMT)}, have the potential to expand the discovery space of the exoplanets from gas giant planets to rocky planets (orbiting M dwarfs) owing to their larger collecting area and greater resolution. Estimating the  polarization-induced aberrations of these
large telescopes is crucial for understanding their magnitude and effect on the high-contrast imaging observations of planets and disks.
The spatially integrated polarization analysis for the TMT and ELT has already been presented by \cite{anche2018analysis,anche2018estimation}, and \cite{de2014instrumental}, respectively, who show that the fold mirrors in the optical configuration are the primary source of polarization effects. In the case of ELT, \cite{de2014instrumental} estimated an instrumental polarization (IP) of 6\% and crosstalk (CT) of 30\% at a wavelength of 0.55 $\mu$m with the primary mirror of ELT considered as a monolith. For the TMT with primary mirror as a monolith, \cite{anche2018analysis} estimates an IP of 4.5\% to 0.6\% and CT of 73\% to 11\% in the wavelength range of 0.4 \textrm{$\mu$m} to 2.583 \textrm{$\mu$m} for field angle and zenith angle equal to zero. Although these results indicate the source and magnitude of polarization introduced by the telescope optics, they focus more on the polarimetric instruments for these telescopes rather than the high-contrast imaging instruments. \cite{yabar2022polarimetric} have also analyzed polarimetric behavior for the segmented mirrors due to aging, dust deposition, and oxidation of the coating. However, a detailed analysis of the polarization aberrations and their effect on the final achievable contrast is required for all three large telescopes using their actual optical configurations, including primary mirror segments. As GSMTs will have a primary mirror of  F/1 compared to the F/2 for the VLTs, we can expect the effect of the polarization aberrations to be more significant compared to the VLTs as the aberrations increase quadratically. 
\par In this series of papers we investigate the effect of the polarization aberrations on the coronagraphic performance and stellar PSF structure incorporating the telescope dynamics (e.g., pointing variations), segment-to-segment variations, coating non-uniformity, missing segments, and wavefront control of the high-order adaptive optics system. In this paper we present the static polarization aberration model and the estimation of polarization aberrations and their effect on coronagraphic performance. 
\par  Section \ref{sec2} provides a brief description of the optical configurations of all three telescopes. The summary of the polarization ray-tracing algorithm and estimation of polarization aberrations is given in  Section \ref{sec3}. The Jones pupil maps evaluated at the exit pupil of the telescope are given in  Section \ref{sec5}. Section  \ref{sec4} shows the diattenuation and retardance variation on all the mirror surfaces for the coatings used in these telescopes. 
The description of the coronagraphs and simulations to estimate the achievable contrast is given in Section \ref{sec6}. Mitigation and calibration strategies are discussed in Section \ref{sec7}. Finally, a summary and conclusions are provided in Section \ref{sec8}.
\section{Description of the optical layout and proposed high-contrast imaging instruments}\label{sec2}
We provide a brief description of the optical layout and the proposed coronagraphic instruments for each telescope. Table \ref{tab:tele-param} provides a summary of the optical design for all three telescopes. 
\begin{table*}[!ht]
\centering
\begin{tabular}{|llll|}
\hline
\multicolumn{1}{|l|}{Telescope   parameters}            & \multicolumn{1}{l|}{ELT}               & \multicolumn{1}{l|}{TMT}                   & GMT                         \\ \hline
\multicolumn{1}{|l|}{Field of View   (Arc minutes)}   & \multicolumn{1}{l|}{10}                & \multicolumn{1}{l|}{15}                    & 20                          \\ \hline
\multicolumn{1}{|l|}{Wavelength of   operation ($\rm \mu$m)}       & \multicolumn{1}{l|}{0.32-25}   & \multicolumn{1}{l|}{0.32-28 }       & 0.32-25              \\ \hline
\multicolumn{4}{|l|}{}                                                                                                                                                    \\ \hline
\multicolumn{1}{|l|}{Primary   mirror-M1}             & \multicolumn{1}{l|}{Elliptical concave}  & \multicolumn{1}{l|}{Hyperbolic   concave}  & Elliptical concave            \\ \hline
\multicolumn{1}{|l|}{Diameter (m)}                        & \multicolumn{1}{l|}{39}                & \multicolumn{1}{l|}{30}                    & 25                          \\ \hline
\multicolumn{1}{|l|}{Conic constant}                  & \multicolumn{1}{l|}{-0.996}            & \multicolumn{1}{l|}{-1.00097}              & -0.998                      \\ \hline
\multicolumn{1}{|l|}{Segment type}                    & \multicolumn{1}{l|}{Quasi Hexagonal}   & \multicolumn{1}{l|}{Quasi Hexagonal}       & Circular                    \\ \hline
\multicolumn{1}{|l|}{Segment size (m)}                    & \multicolumn{1}{l|}{1.45}              & \multicolumn{1}{l|}{1.44}                 & 8.365                      \\ \hline
\multicolumn{1}{|l|}{Segment gap}                     & \multicolumn{1}{l|}{4mm}               & \multicolumn{1}{l|}{2.5mm}                 & 0.345m                      \\ \hline
\multicolumn{1}{|l|}{Number of   segments}            & \multicolumn{1}{l|}{798}               & \multicolumn{1}{l|}{492}                   & 7                           \\ \hline
\multicolumn{4}{|l|}{}                                                                                                                                                    \\ \hline
\multicolumn{1}{|l|}{Secondary   mirror-M2}           & \multicolumn{1}{l|}{Hyperbolic convex}            & \multicolumn{1}{l|}{Hyperbolic convex}                & Elliptical concave                     \\ \hline
\multicolumn{1}{|l|}{Diameter (m)}                        & \multicolumn{1}{l|}{4.1}               & \multicolumn{1}{l|}{3.1}                   & 1.05                        \\ \hline
\multicolumn{1}{|l|}{Conic constant}                  & \multicolumn{1}{l|}{-2.28962}          & \multicolumn{1}{l|}{-1.381}                & -0.716927274                \\ \hline
\multicolumn{1}{|l|}{Segments}                        & \multicolumn{1}{l|}{No}                & \multicolumn{1}{l|}{No}                    & Yes                         \\ \hline
\multicolumn{1}{|l|}{Number of   Segments}            & \multicolumn{1}{l|}{NA}                & \multicolumn{1}{l|}{NA}                    & 7                           \\ \hline
\multicolumn{1}{|l|}{Active/Adaptive control}                 & \multicolumn{1}{l|}{Yes}               & \multicolumn{1}{l|}{No}                    & Yes                         \\ \hline
\multicolumn{4}{|l|}{}                                                                                                                                                    \\ \hline
\multicolumn{1}{|l|}{Tertiary   mirror-M3}            & \multicolumn{1}{l|}{Circular Concave}           & \multicolumn{1}{l|}{Flat}       &  Flat               \\ \hline
\multicolumn{1}{|l|}{Diameter (m)}                        & \multicolumn{1}{l|}{4}                 & \multicolumn{1}{l|}{3.5$\times$2.5}               & 0.3                          \\ \hline
\multicolumn{1}{|l|}{Inclined angle(\textdegree)}                  & \multicolumn{1}{l|}{0}              & \multicolumn{1}{l|}{45}                    & 45                          \\ \hline
\multicolumn{4}{|l|}{}                                                                                                                                                    \\ \hline
\multicolumn{1}{|l|}{Quaternary   mirror-M4}          & \multicolumn{1}{l|}{Flat}              & \multicolumn{1}{l|}{NA}                    & NA                          \\ \hline
\multicolumn{1}{|l|}{Diameter}                        & \multicolumn{1}{l|}{2.38$\times$2.34}         & \multicolumn{1}{l|}{}                      &                             \\ \hline
\multicolumn{1}{|l|}{Inclined angle(\textdegree)}                  & \multicolumn{1}{l|}{7.25}             & \multicolumn{1}{l|}{}                      &                             \\ \hline
\multicolumn{4}{|l|}{}                                                                                                                                                    \\ \hline
\multicolumn{1}{|l|}{Fifth   Mirror-M5}               & \multicolumn{1}{l|}{Flat}              & \multicolumn{1}{l|}{NA}                    & NA                          \\ \hline
\multicolumn{1}{|l|}{Diameter}                        & \multicolumn{1}{l|}{2.6$\times$2.1}           & \multicolumn{1}{l|}{}                      &                             \\ \hline
\multicolumn{1}{|l|}{Inclined angle (\textdegree)}                  & \multicolumn{1}{l|}{37.25}             & \multicolumn{1}{l|}{}                      &                             \\ \hline
\multicolumn{4}{|l|}{}                                                                                                                                                    \\ \hline
\multicolumn{1}{|l|}{Proposed coating}& \multicolumn{1}{l|}{Protected Silver}& \multicolumn{1}{l|}{Protected Silver}&
\multicolumn{1}{|l|}{Bare Aluminum} \\ \hline
\multicolumn{4}{|l|}{}  \\ \hline
\multicolumn{1}{|l|}{Final focal ratio}                  & \multicolumn{1}{l|}{f/17.48}             & \multicolumn{1}{l|}{f/15}  & f/8.2 \\ \hline
\multicolumn{1}{|l|}{Instrument   mounting locations} & \multicolumn{1}{l|}{Nasmyth platforms} & \multicolumn{1}{l|}{Nasmyth platforms}     & Direct Gregorian            \\ \hline
\multicolumn{1}{|l|}{}                                & \multicolumn{1}{l|}{A and B}           & \multicolumn{1}{l|}{A and B}               & Folded port                 \\ \hline
\multicolumn{1}{|l|}{}                                & \multicolumn{1}{l|}{}                  & \multicolumn{1}{l|}{}                      & Auxiliary ports             \\ \hline
\end{tabular}
\caption{Telescope parameters for the three telescopes \citep{cayrel2012elt,nelson2006tmt,bernstein2014overview}}
\label{tab:tele-param}
\end{table*}
\subsection{Extremely Large Telescope} The ELT, with a 39m diameter segmented primary mirror, is currently being constructed on the top of Cerro Armazones in the Atacama Desert of northern Chile, and is expected to have its first light in 2027. The optical layout of the telescope has five mirrors with an intermediate focus, as shown in Figure \ref{opl-gsmt}. The Multi-AO Imaging Camera for Deep Observations (\href{https://elt.eso.org/instrument/MICADO/} {MICADO}) will have  a multi-conjugate adaptive optics (MCAO) module (as a part of the second-generation capability) feeding the coronagraphic imager in the wavelength range of 0.8–2.4$\rm \mu$m \citep{davies2021micado}. 
One of the first-generation instruments of ELT, the Mid-Infrared ELT Imager and Spectrograph (\href{https://elt.eso.org/instrument/METIS/}{METIS}) will also have a coronagraphic imaging capability with a single conjugate adaptive optics system in the wavelength range of 3-13$\mu$m with a field of view (FOV) of 10 arcsec \citep{brandlmetis2021}. The instrument with eXtreme adaptive optics (XAO), the Planetary Camera, and Spectrograph (PCS), will perform coronagraphy with spectroscopic and polarimetric capabilities. It is being designed to support both blue and red channels expected to achieve a post-processed contrast of $\rm 10^{-8}$ at 15 mas and $\rm 10^{-9}$ at 100 mas angular separation from the star \citep{kasper2021}. The required inner working angle (IWA) of 15 mas is between 1 and 3$\lambda/D$ where a raw contrast of $10^{-5}$ to $10^{-4}$ is expected \citep{kasper2021, nousiainen2022toward}.
\subsection{Thirty Meter Telescope}
The TMT, with a 30m diameter segmented primary mirror, has been proposed to be built on Mauna Kea, Hawaii, USA, and is expected to have its first light in the early 2030s. The telescope is a Ritchey Chr\'etien with a fold mirror (shown in Figure \ref{opl-gsmt}) to direct light to different instruments on the Nasmyth platform. The Planetary Systems Imager (\href{https://psi.astro.ucla.edu/}{PSI}) is a proposed second-generation instrument with coronagraphic capability in both blue (0.6-1.8$\rm \mu$m) and red channels (2-5$\rm \mu$m and 8-13$\rm \mu$m) and is expected to achieve a post-processed contrast of $\rm 10^{-8} $ at $\rm 2\lambda/D$. This requires a raw contrast of $10^{-5}$ at 1 to 2$\lambda/D$ \citep{fitzgerald2019planetary}. The current design of PSI supports polarimetric imaging and fiber-fed high-resolution spectroscopy (R$\sim$100,000) \citep{fitzgerald2022planetary}.
\subsection{Giant Magellan Telescope}
\begin{figure*}[!ht]
\centering
\includegraphics[width=1\textwidth]{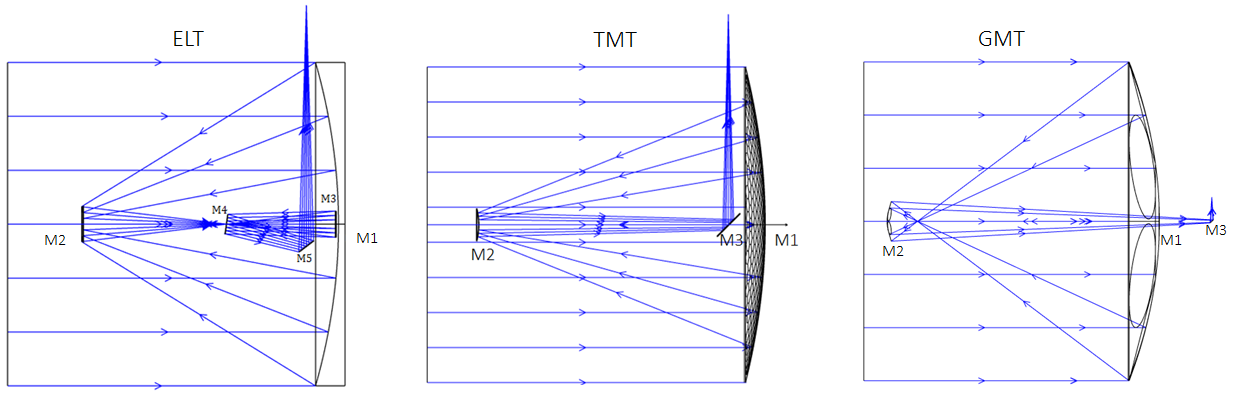}
\caption{Optical layout of the telescopes from Zemax\textsuperscript{\textregistered} for   the three telescopes}
\label{opl-gsmt}
\end{figure*}

The 25m diameter GMT is being built on Las Campanas Peak at the southern edge of Chile’s Atacama Desert and is expected to have its first light in 2029. It is designed to support imaging at both the Gregorian and Nasmyth focus. The optical layout of GMT is shown in Figure \ref{opl-gsmt} with the fold mirror. The Giant Magellan Adaptive Optics eXtreme (GMagAO-X) instrument is one of the proposed first-generation instruments with coronagraphic capabilities. It will provide wavelength coverage from g band to K band with broadband imaging and a fiber-fed integral field unit (IFU) spectrograph with R$\sim$ 100,000. It is expected to achieve a post-processed contrast $10^{-8}$ for an eighth magnitude star at a wavelength of 800nm, which requires a raw contrast of $10^{-5}$ at 1 to 2$\lambda/D$ \citep{males2022conceptual}.
\subsection{Optical design of telescopes in Zemax}
We used the exact segment prescription for the primary mirror of TMT and GMT. In the case of ELT, we   simulated a monolithic primary mirror and incorporated the segment aperture mask after performing a ray trace (explained in the next section) in Zemax\textsuperscript{\textregistered} version 2022. For TMT and GMT, the telescope is designed in the mixed mode (combination of sequential and non-sequential mode), where the segments are added as non-sequential components. First, each segment is specified using a user-defined aperture (UDA) file containing the positions in the global coordinate system (e.g., vertices of the hexagonal segments for TMT). Then, the secondary and Nasmyth mirrors of the telescope are added in sequential mode.
\section{Polarization ray tracing algorithm} \label{sec3}
Modeling the effects of polarization in optical systems is typically conducted in the Jones or Mueller formalism. Jones calculus treats polarizing optical elements as 2x2 complex matrices and traces the evolution of the complex electric field. The Mueller calculus is a 4x4 matrix formalism that propagates the incoherent power in three polarization states and the degree of polarization. Because modern detectors can only measure the irradiance from optical fields rather than the fields themselves, the Mueller calculus represents the irradiance distribution that can be calculated from a system. However, we require the Jones representation to trace the influence on the complex amplitude to simulate the diffraction effects in high-contrast imaging instruments. Therefore, we employ polarization ray tracing (PRT) to calculate the total Jones matrix experienced by every ray propagating through the optical system. These Jones matrices can be converted to Mueller matrices \citep{chipman2018polarized} to retrieve the final intensity in response to an unpolarized star. 
\par
\begin{figure}[!ht]
\centering
\includegraphics[trim={4cm 15cm 6cm 7cm},clip,width=\columnwidth]{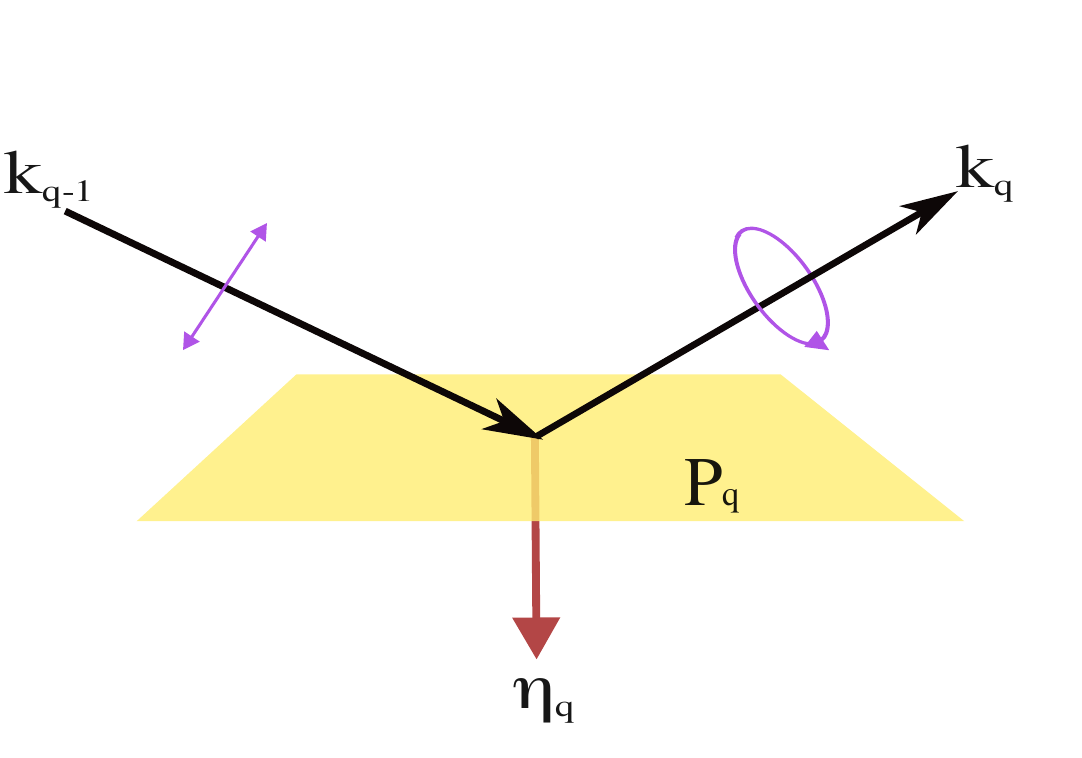}
\caption{Diagram illustrating a typical polarization ray tracing interaction in reflection with the $q$-th surface in the optical system. The wave vector before the surface $\mathbf{k}_{q-1}$ is incident on the surface with a linear polarization shown in purple. The angle of reflection is determined by the angle of incidence with respect to the surface normal $\mathbf{\eta}_{q}$, which points into the surface to maintain a right-handed coordinate system. Next, the polarization state is transformed by the surfaces PRT matrix $\mathbf{P}_{q}$ and propagated along the existent wave vector $\mathbf{k}_{q}$. Here $\mathbf{P}_{q}$ is a partial retarder, so the polarization state becomes elliptical.}
    \label{fig:prt_interaction}
\end{figure}

Polarization ray tracing is a method of computing how the polarization state transforms through an optical system (illustrated in Figure \ref{fig:prt_interaction}). The polarization state is propagated along geometrical ray paths in global coordinates through the optical system. When a ray encounters the surface, the polarization state is rotated into the local coordinate system of the ray-surface interaction. The orthogonal transformation matrices ($\mathbf{O_{in/out}}$) are constructed from the eigenpolarizations of the local surface and the surface normal ($\eta$). In the local coordinate system the Fresnel reflection coefficients are computed for each eigenpolarization and organized into a diagonal matrix $\mathbf{J_{q}}$. The matrix that encodes the orthogonal transformations and Fresnel reflection coefficients is called the PRT Matrix, $\mathbf{P}_{q}$, and is computed using equation \ref{eq:PRT_matrix}:
\begin{equation}
\mathbf{P_{q}} = \mathbf{O_{out,q}} \mathbf{J_{q}} \mathbf{O_{in,q}^{-1}}
\label{eq:PRT_matrix}  
.\end{equation}
The influence of the entire optical system is accomplished through a matrix multiplication of the $\mathbf{P}_{q}$ matrices for $Q$ optical elements:
\begin{align}
\mathbf{P_{tot}} = \prod_{q=0}^{Q} \mathbf{P_{q}}.
\label{eq:PRT_total}
\end{align}

The resultant total PRT matrix $\mathbf{P_{tot}}$ represents the total three-dimensional transformation of the polarization of light from the optical system in global coordinates. To transform this matrix into something useful for diffraction models of coronagraphs, we must compute the Jones pupil. The Jones pupil is the result of an orthogonal transformation of the $\mathbf{P_{tot}}$ matrix into the local coordinates of the exit pupil. To accomplish this transformation, we derive the basis vectors of the entrance pupil and exit pupil and organize them into orthogonal transformation matrices ($\mathbf{O}_{EP}$ and $\mathbf{O}_{XP}$, respectively). The Jones pupil is computed by solving for $\mathbf{J_{tot}}$ in equation \ref{eq:PRT_matrix} using $\mathbf{O}_{EP}$ as $\mathbf{O}_{in}$ and $\mathbf{O}_{XP}$ as $\mathbf{O}_{out}$:
\begin{equation}
\mathbf{J_{tot}} = \mathbf{O}_{XP}^{-1} \mathbf{P_{tot}} \mathbf{O}_{EP}
\label{eq:Jones_Pupil}
.\end{equation}
The $\mathbf{J_{tot}}$ matrix should be zero value in the last row and column, except for the element on the diagonal, which should be unity. This indicates that the matrix only operates on the components of the Jones vector orthogonal to propagation, which is where the electric field is located.
There are several bases to choose from to derive the orthogonal transformation matrices for the entrance and exit pupils \citep{chipman2018polarized}: the \emph{s,p,k} basis, dipole basis, and  double-pole basis. We use the double-pole basis because of its insensitivity to polarization singularities. For an instructive description of the polarization ray tracing algorithm and double-pole coordinate system, we refer  to Chapters 10 and 11 of \cite{chipman2018polarized}.
\par In this work we perform ray tracing for an array of 256$\times$ 256 rays in Zemax\textsuperscript{\textregistered} using the Python ZOS-API, and generate Jones matrices using a polarization ray tracing module \citep[based on][]{chipman2018polarized} developed in Python\footnote{https://github.com/Jashcraf/poke} \citep{Ashcraft_poke_2022}. The ray trace in Zemax\textsuperscript{\textregistered} provides the incident direction cosine and surface normal at each ray intercept and the corresponding angle of incidence. We calculate the direction cosines of the reflected ray and local \textit{s,p} eigenpolarizations at each mirror surface which form the $\mathbf{O_{out,q}}$ and $\mathbf{O_{in,q}^{-1}}$ matrices. 
\subsection{Angle of incidence (AOI) on the mirror surfaces}
\begin{figure}[!ht]
\includegraphics[width=\columnwidth]{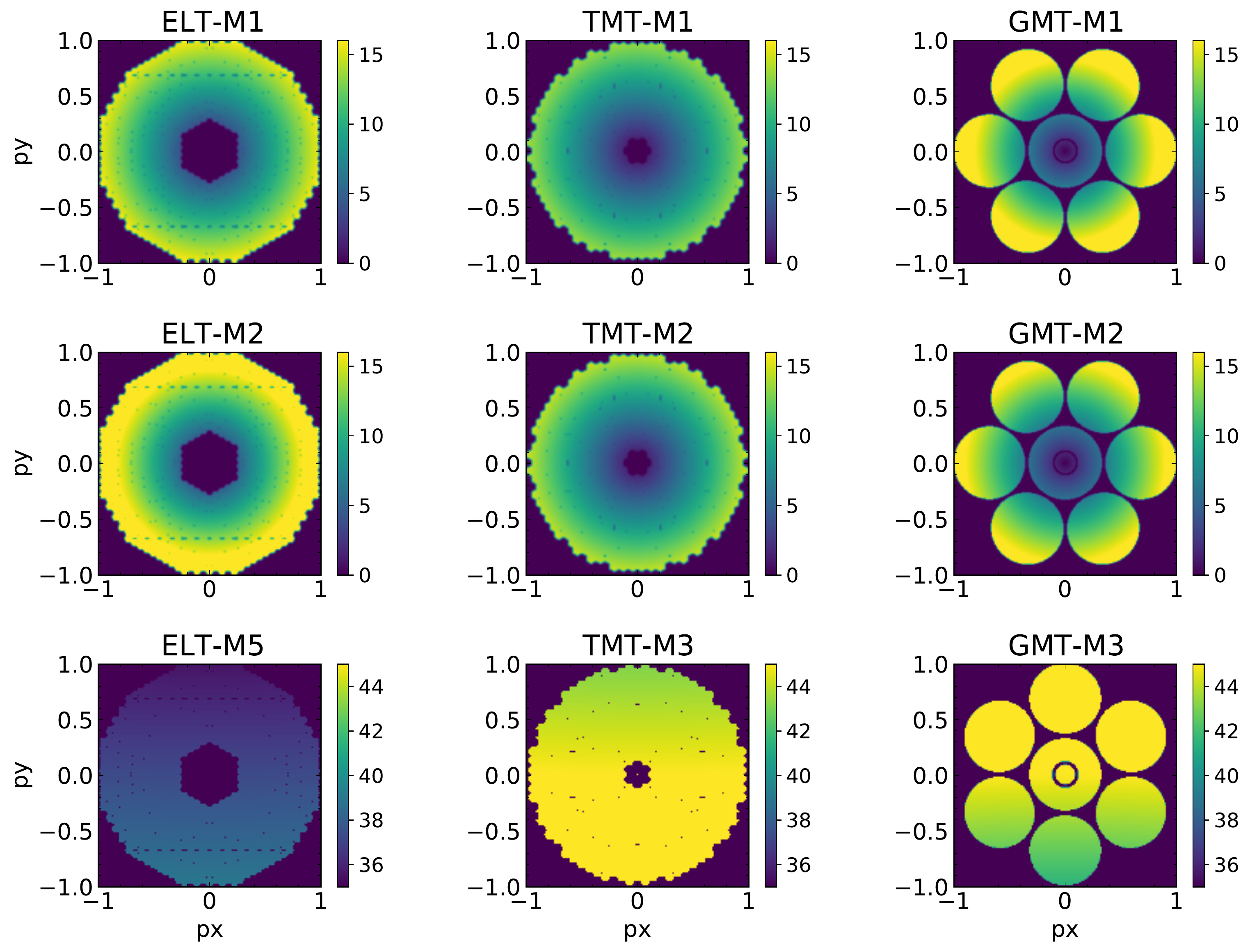}
\caption{Variation of incident angles (\textdegree) for different mirrors in the three telescopes. Mirror M5 is shown for the ELT, which directs the light to the different instruments on the Nasmyth platform (analogous to M3 in TMT and GMT). \textit{px} {and} \textit{py} corresponds to the normalized entrance pupil coordinates.}
\label{incident-angles}
\end{figure}

The AOI is obtained by performing ray trace in Zemax\textsuperscript{\textregistered} for all the mirror surfaces. The maximum AOI for the primary mirror is 16.26\textdegree, 14.01\textdegree, and 19.46\textdegree$~$ for ELT, TMT, and GMT, respectively. The secondary mirror has a similar range of AOI for all three telescopes. In the case of ELT, the maximum AOI on M3 and M4 is estimated to be 2.56\textdegree$~$and 9.49\textdegree,$~$ respectively. For the final fold mirror, the AOI  varies between 35.5\textdegree \ and  39.017\textdegree$~$for ELT, 42.98\textdegree\  and 46.89\textdegree$~$for TMT, and 41.5\textdegree\ and  48.5\textdegree$~$for GMT, respectively. The AOI obtained from ray trace in Zemax\textsuperscript{\textregistered} is shown for   the mirror surfaces for TMT and GMT in Figure \ref{incident-angles}, ignoring the spider structure. In the case of ELT, we show the mirror M5, which feeds the light to different instruments on the Nasmyth platform. The incident angles on M3 and M4 of ELT vary in the range 0-2.56\textdegree$~$(center to edge) and 6.01-9.49\textdegree$~$(along the y-axis), respectively.
\par 
The polarization introduced from the mirror surface increases with the AOI on the mirror surface as the difference between the reflection coefficients $r_p$ and $r_s$ increases \citep{giro2003polarization}. The primary and secondary mirrors have a smaller angle of incidences $\sim$ 15-20\textdegree$~$ and introduce nonzero ($\sim$ 0.1\%) instrumental polarization and crosstalk due to their segmented apertures. On the other hand, the fold mirrors with the highest angle of incidence contribute to a significant part (1-2\% in $V$ band) of the instrumental polarization and polarization crosstalk. In the context of polarization aberrations, M1 and M2 of TMT and GMT contribute to the diattenuation- and retardance-defocus, and M3 gives rise to diattenuation piston and tilt, and retardance piston and tilt. For ELT, M1, M2, and M3 contribute to the diattenuation- and retardance-defocus, and M4 and M5 give rise to diattenuation piston and tilt, and retardance piston and tilt.
\subsection{Computation of Fresnel reflection coefficients from thin films}
\begin{figure}
\includegraphics[width=1\linewidth]{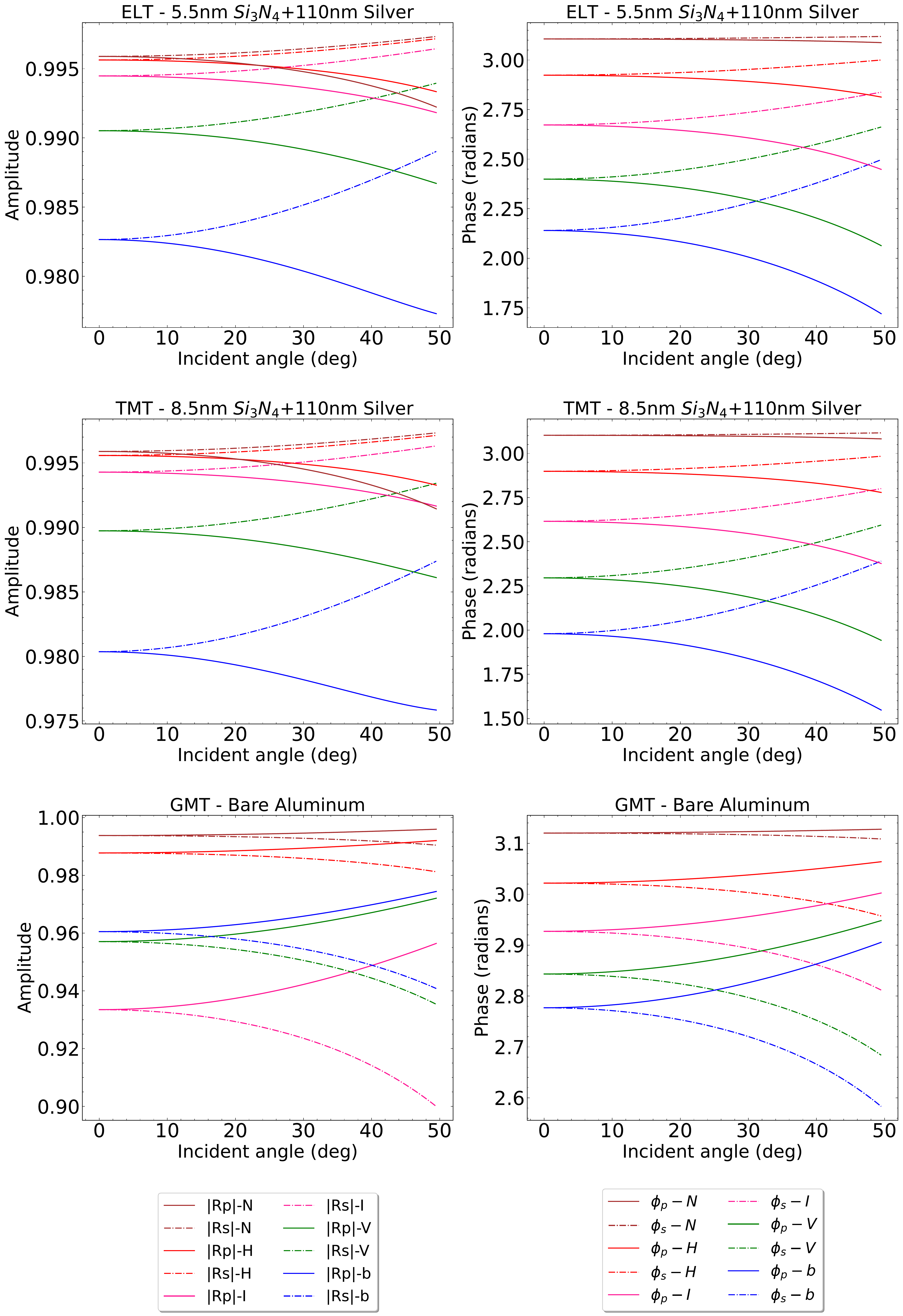}
\caption{Variation of amplitude reflection coefficients in \textit{p} and \textit{s} with the angle of incidence   for different astronomical filter bands. ELT and TMT mirrors will have a Gemini-like coating with silver as the main reflective layer and Si3N4 as the protective layer, whereas GMT will have a bare aluminum coating.}
\label{amp-ref}
\end{figure}

The ELT and TMT observatory mirrors are overcoated with a dielectric material to protect the reflective silver mirror coating. 
This dielectric alters the effective Fresnel reflection coefficient of the mirrors, which was considered in our model. 
To simulate the effective Fresnel reflection coefficient, we employ the method outlined in \citet{macleod2010thin}, which is reproduced below. The effective Fresnel reflection coefficient is derived from the characteristic matrix of the thin film. 
This matrix is given by
\begin{equation}
\begin{pmatrix}
B \\
C \\
\end{pmatrix}
=
\prod_{q=1}^{N-1} 
\begin{pmatrix}
cos(\delta_{q}) & isin(\delta_{q})/\eta_{q} \\
\eta_{q}sin(\delta_{q}) & cos(\delta_{q}) \\
\end{pmatrix}
\begin{pmatrix}
1 \\
\eta_{N}
\end{pmatrix},
\label{eq:char_matrix1}
\end{equation}
where $\delta_{q}$ is the phase thickness of the $q$-th film, given by 
\begin{equation}
\delta_{q} = \frac{2\pi d_{q} cos(\theta_{q})}{\lambda}.
\label{eq:phase_thickness}
\end{equation}
Here, $\eta_{q}$ is the characteristic admittance of the $q$-th film, which is defined as 
\begin{gather}
\eta_{q,s} = n_{q} cos(\theta_{q}), \\ \eta_{q,p} = n_{q} / \cos(\theta_{q}),
\label{eq:admittance}
\end{gather}
and $\theta_{q}$ is the angle of the wave vector in the $q$-th film. Solving the characteristic matrix for the $B$ and $C$ coefficient yields the effective reflection coefficient ($r_{eff}$) of the thin film stack
\begin{equation}
r_{eff} = \frac{\eta_{0}B - C}{\eta_{0}B + C},
\end{equation}
where $\eta_{0}$ is the characteristic admittance of the medium (free space). This formalism assumes that the substrate is a solid substrate of the reflective material (silver, aluminum) with a single dielectric coating as the skin depth of metals at optical frequencies is typically a fraction of the wavelength \citep{BornWolf:1999:Book}\footnote{The calculation is given in Ch 14.2 Equation 22. For example, the energy density of a $\lambda=500nm$ wave falls to 1/e at ~12nm in Ag and 6.5nm in Al}. The resulting characteristic matrix is
\begin{equation}
\begin{pmatrix}
B \\
C \\
\end{pmatrix}
=
\begin{pmatrix}
cos(\delta_{SiN}) & isin(\delta_{SiN})/\eta_{SiN} \\
\eta_{SiN}sin(\delta_{SiN}) & cos(\delta_{SiN}) \\
\end{pmatrix}
\begin{pmatrix}
1 \\
\eta_{Ag}
\end{pmatrix},
\label{eq:char_matrix}
\end{equation}
where \textit{SiN} corresponds to $\rm Si_3N_4$ as the dielectric layer in our model. The ELT and TMT mirrors will have Gemini-like four-layer coating. The ELT coating has 60$\angstrom$  thick $\rm NiCrN_x$ on the Zerodor substrate followed by 1100$\angstrom$ of silver,  3$\angstrom$ of $\rm NiCrN_x$, and finally 55$\angstrom$ thick aluminum-doped $\rm Si_3N_4$ \citep{schotsaert2020coating}. The TMT will have 65$\angstrom$ thick $\rm NiCrN_x$ on the Zerodor substrate followed by 1100$\angstrom$ of silver,  6$\angstrom$ of $\rm NiCrN_x$, and finally 85$\angstrom$ thick $\rm Si_3N_4$ as the top layer \citep{anche2018analysis}. The GMT mirrors will be coated with bare aluminum. In our analysis we do not consider $\rm NiCrN_x$ for the ELT and TMT mirrors due to the unavailability of the refractive index information and the influence of the aluminum oxide layer for the GMT mirrors \citep{van2009polarization}.
However, we expect the impact of $\rm NiCrN_x$ to be small due to the small phase thickness of this layer. Figure \ref{amp-ref} shows the amplitude and phase of the reflection coefficients for five astronomical filter bands (\textit{b}-\textit{N}) for all three coatings. The amplitude of the reflection coefficients is $>$ 0.975 for TMT and ELT, as the coatings have been optimized for higher reflectivity. In contrast, the reflection coefficient phase varies by almost 1 radian over these wavelengths. For GMT, the amplitude of reflection coefficients varies from 0.90 to 0.98, and the phase varies on the order of 0.3 radians over the wavelengths. The difference between the amplitude ($|R_p|-|R_s|$) and phase ($|\phi_p|-|\phi_s|$) reduces from blue to red wavelengths, and predicts similar behavior in the polarization aberrations. 
\section{Jones pupil maps}
\label{sec5}
\begin{figure*}
\centering
\includegraphics[width=1\textwidth]{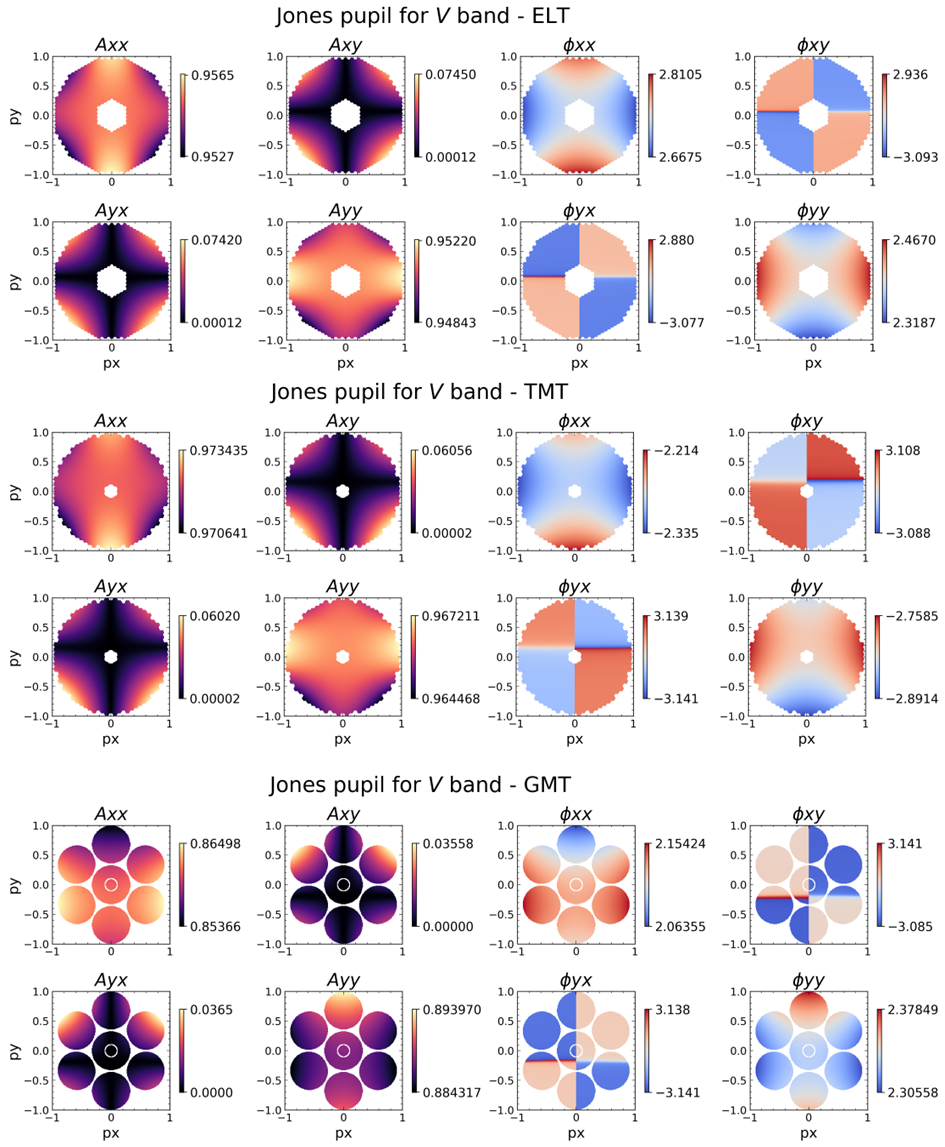}
\caption{Jones pupil maps for the GSMTs shown for $V$-band filter estimated at the telescope's exit pupil. \textit{Axx} and \textit{Ayy} show the amplitudes for \textit{X} and \textit{Y} polarized light, respectively, and \textit{Axy} and \textit{Ayx} are the cross-coupled components. $\phi xx$ and $\phi yy$ show the phase in radians for $X$ and $Y$ polarized light and  $\phi xy$ and $\phi yx$ are the cross-coupled components}
\label{jones-GSMT}
\end{figure*}

As explained in section \ref{sec3}, the polarization aberrations can be expressed in terms of the Jones pupil map (Jones matrices as a function of object and pupil coordinates) obtained using the PRT through the telescope. The Jones pupil map shows how incident X and Y electric fields are manipulated through the system. Ideally, in the absence of polarization aberrations, the Jones matrix obtained at the exit pupil will be an identity matrix, and the integrated on-axis PSF will be unpolarized. However, due to the complex refractive indices of the coating and the curvature of the mirrors, it deviates from the ideal scenario. Figure \ref{jones-GSMT} shows the Jones pupil map generated at the exit pupil of each of the telescopes in the $V$ band. \textit{Axx} and \textit{Ayy} show the transmission for \textit{X} and \textit{Y} polarized light, respectively. \textit{Axy} and \textit{Ayx} show the amplitude of \textit{Y}-polarized light converted to \textit{X}-polarized light and vice versa. A shifted Maltese cross pattern is seen in \textit{Axy} and \textit{Ayx} for all three telescopes, indicating crosstalk between the  X and Y polarization. The crosstalk terms \textit{Axy} and \textit{Ayx} are highly apodized for all three telescopes, and their amplitude is highest (7.45\%) for the ELT and lowest for the GMT (3.5\%). The difference in the \textit{Axx} and \textit{Ayy} terms is $\sim$ 0.6\% for ELT and TMT and $\sim$ 3\% for GMT, which can be attributed to the difference between the amplitude of reflection coefficients.

The panels $\phi xx$ and $\phi yy$ show the phase shift for $X$ and $Y$ polarized light, respectively, and $\phi xy$ and $\phi yx$ show the phase shift of \textit{Y}-polarized light converted to \textit{X}-polarized light and vice versa. $\phi xx$ and $\phi yy$  vary over the pupil for all three telescopes showing differential astigmatism between XX and YY, which manifests as retardance defocus and tilt. $\phi xx$ and $\phi yy$ vary $\sim$ 12nm, 10nm, and 7.5nm over the pupil for all three telescopes. Comparing the amplitude apodization and phase variation in the Jones pupil, it can be seen that polarization aberrations in ELT will have a larger impact on the coronagraphic performance than the other two GSMTs.
These Jones pupils of the three telescopes are fit using the six analytical polarization aberration terms as shown in \cite{breckinridge2015polarization}, and the coefficients of the aberrations are provided in Appendix B for all the filter bands.
\section{Comparison of coatings: Diattenuation and retardance}
\label{sec4}
\begin{figure}
    \centering
    \includegraphics[width=1\columnwidth]{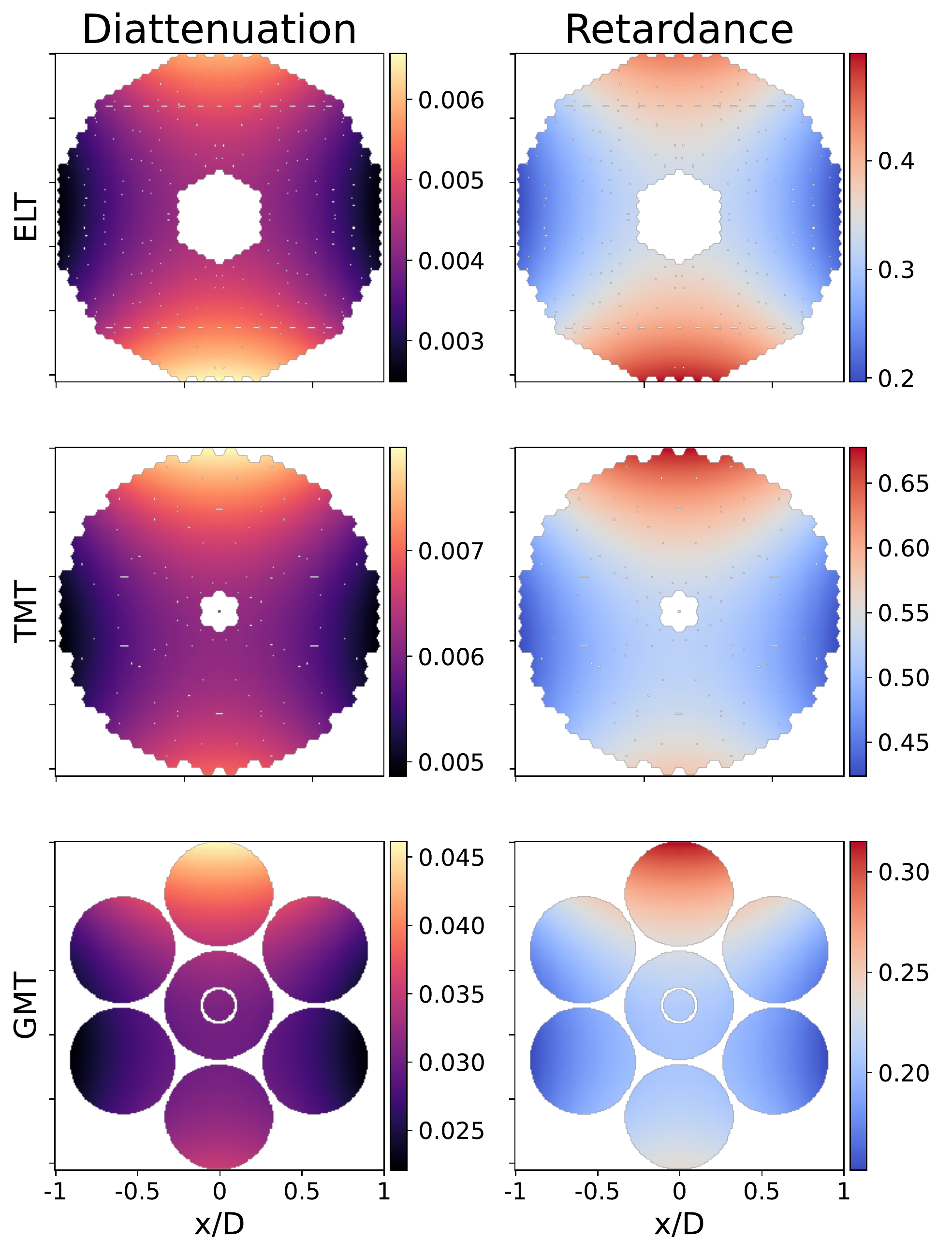}
    \caption{Comparison of the total diattenuation (left) and retardance (right) of the E-ELT (top), TMT (middle), and GMT (bottom) in the V band for X-polarization. These data reveal the tilted astigmatic pattern characteristic of the polarization aberrations expected for Cassegrain telescopes with fold mirrors. Y-polarized light experiences a similar pattern rotated by 90 degrees. In addition, these data reveal that the protected silver mirrors experience greater retardance and lower diattenuation than the bare-aluminum GMT.}
    \label{fig:total_D_R}
\end{figure}

To study polarization aberrations, it is convenient to decompose the Jones pupil into diattenuation and retardance. For homogeneous Jones matrices, the diattenuation ($D$) and retardance ($\delta$) are computed from the eigenvalues of the Jones matrix ($\xi_x,\xi_y$)
\begin{equation}
    D = \frac{|\xi_x|^{2} - |\xi_y|^{2}}{|\xi_x|^{2} + |\xi_y|^{2}}
,\end{equation}
\begin{equation}
    \delta = |\angle \xi_x - \angle \xi_y|
,\end{equation}
where $\angle$ is the angle operator and $\xi_x$, and $\xi_y$ are the eigenvalues of maximum and minimum polarization aberration, respectively. This operation is performed on the Jones pupil to examine the diattenuation and retardance expressed in the local basis vectors of the exit pupil, which serves as the entrance pupil of a coronagraph. We examine the spatial variation and performance versus astronomical band to assess how each polarization aberration influences the coronagraphic performance.
Figure \ref{fig:total_D_R} plots the diattenuation and retardance for X polarization across the Jones pupil in the V band to reveal a shifted astigmatic pattern that is characteristic of Cassegrain-type telescopes with a fold mirror \citep{breckinridge2015polarization}. The protected silver coating used by the TMT and ELT is less absorbing in the $V$ band when compared to the aluminum coating used by the GMT. 
However, due to the protective dielectric coating, the ELT and TMT experience greater retardance than GMT. Between the TMT and ELT, the ELT has lower peak-to-valley polarization aberration, which can be credited to the inclined fold mirror at 37.5\textdegree, which causes diattenuation and retardance tilt. Figure \ref{incident-angles} shows that the primary and secondary mirrors have comparable angles of incidence. The TMT tertiary mirror imposes significantly greater polarization aberration due to the 45\textdegree\ angle of incidence. The ELT employs two separate flats with lower angles of incidence than the TMT  tertiary (see Table \ref{tab:tele-param}). The sum of the contributions to the polarization aberration of the two mirrors is less than that of a single mirror with a higher angle of incidence, resulting in lower peak-to-valley diattenuation and retardance than the TMT. 
\begin{figure*}
\centering
\includegraphics[width=1.0\linewidth,trim=0 0 0 0]{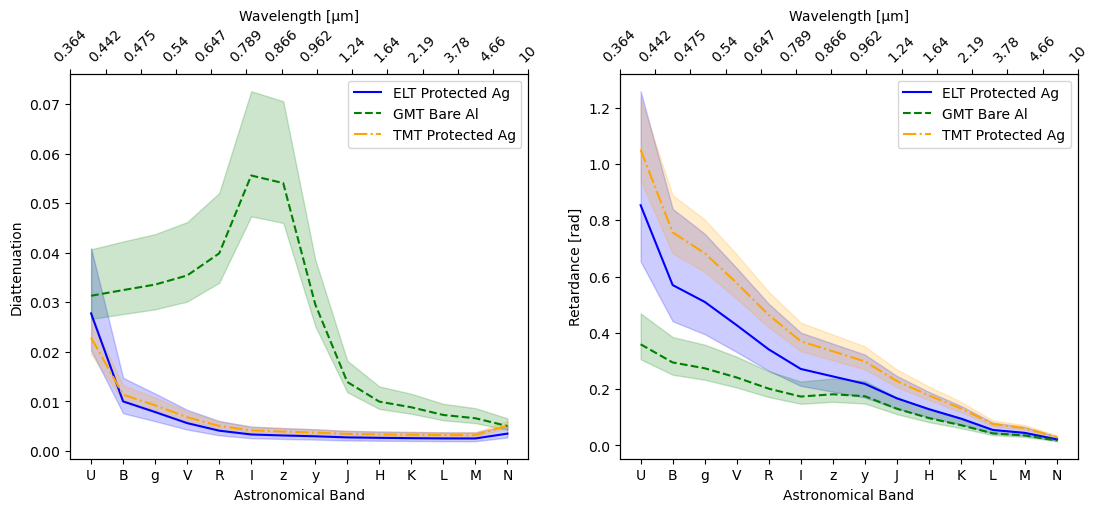}
\caption{Total diattenuation and retardance for each astronomical band. Each solid line represents the mean value of the total diattenuation and retardance. The shaded areas around these lines represent the maximum and minimum absolute values of total diattenuation and retardance. The total diattenuation and retardance decrease with wavelength, as expected, except for the GMT diattenuation due to aluminum's high visible and near-infrared absorption.}
\label{fig:total_D_R_vsband}
\end{figure*}
To examine the performance versus the astronomical band, we compute the minimum, maximum, and mean values of the absolute diattenuation and retardance for each band. These data are plotted in Figure \ref{fig:total_D_R_vsband} and substantiate the trends discussed earlier. The GMT diattenuation is the highest, particularly near the high absorption band in the visible and near-infrared (V-y band), followed by TMT and then ELT. The TMT has the greatest retardance, followed by ELT and then  GMT. The overall polarization aberration (implicitly represented by the shaded regions in Figure \ref{fig:total_D_R_vsband}) tends to decrease with an increase in wavelength.

As an indicator of coronagraphic performance, retardance represents a polarization-dependent phase aberration that will directly shape the point-spread function supplied to the coronagraph. On the other hand, diattenuation represents a polarization-dependent amplitude smoothing rather than explicitly changing the distribution of the PSF. Therefore, we expect the telescopes with high retardance to have worse coronagraphic performance, and diattenuation will be of lesser consequence. To accurately understand the exact relationship of polarization aberrations to coronagraphic performance, we must inject a Jones pupil into a diffraction model of the coronagraph.
\section{Effect on the achievable contrast}
\label{sec6}
\begin{figure*}
\centering
\includegraphics[width=1.0\textwidth]{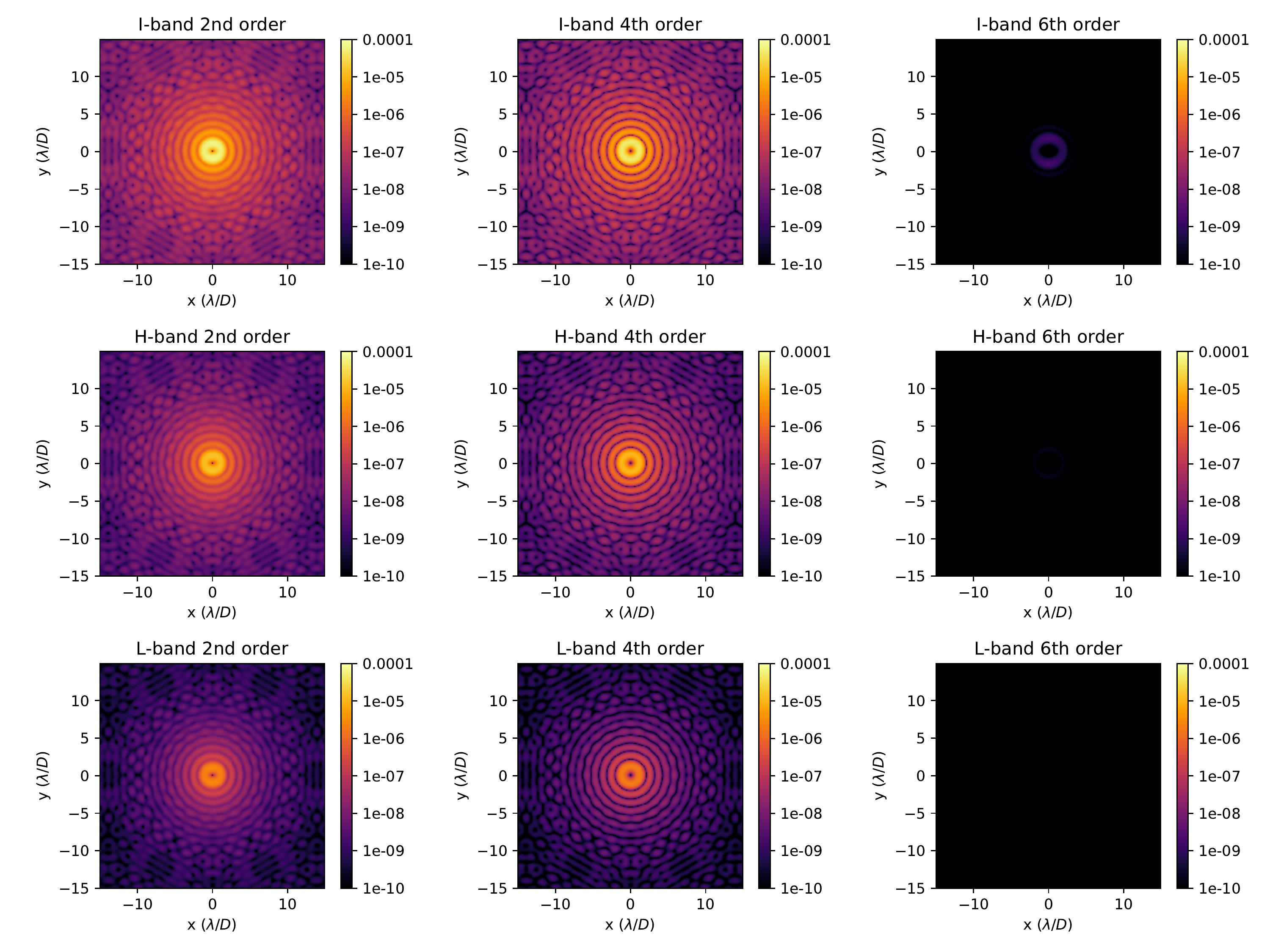}
\caption{Stellar residuals for different wavelengths and coronagraphs for the TMT. The residuals are shown for I band (top), H band (middle), and L band (bottom). The coronagraphs are a second- (left), fourth- (middle), and sixth-order (right) coronagraph. These images show that the stellar residuals is mainly defocus.}
\label{fig:coranagraphic_residuals_TMT}
\end{figure*}

Any differential aberration between the X- and Y-polarization states will leak through a coronagraph because only the common aberration can be compensated. The Jones pupils from the ray trace of the GSMTs are used as an input to our high-contrast imaging simulations with the High-Contrast Imaging in Python (HCIPy) module \citep{por2018hcipy}. The Jones pupils have to be post-processed before propagating through the coronagraphic simulations because they contain empty areas due to the finite ray sampling. The empty pixels are interpolated with a local second-order polynomial in x and y (1, $x$, $y$, $xy$, $x^2$, $y^2$). The local interpolator uses a 5x5 pixel area around each empty pixel to determine the local polynomial coefficients. The 5x5 area is the minimum area size with the empty pixel in the center and enough pixels to fit a second-order polynomial. A 3x3 area creates very strong artifacts at the  edge pixels where there are not enough pixels to constrain the solution. Higher-order  interpolation or larger pixel areas did not significantly change the results. We kept the interpolation scheme at 5x5 to make it as local as possible. The interpolation is done in real and imaginary space, which are continuous because amplitude and phase space are not due to phase jumps (see Figure \ref{jones-GSMT}).

The now interpolated pupils still contain aberrations common between the X and Y states. An AO system cannot distinguish between phase aberrations from the atmosphere or the telescope from polarization aberrations when the system is observing a star. Therefore, the AO system will compensate for all common polarization aberrations between X and Y. This is incorporated in our simulations by taking the average between the phase of the \textit{xx} and \textit{yy} elements of the Jones pupil. The input into our simulations is assumed to be an unpolarized star. HCIPy uses the Stokes vector to generate several fully polarized electric fields that add up to match the input Stokes vector. Each fully polarized electric field (e.g., an X field or Y field) can be propagated through the entire optical system.  
The impact of the aberrations on the contrast is estimated by propagating the electric fields through a set of perfect coronagraphs (PCs) \citep{cavarroc2006coronagraph, guyon2006coronagraph}. These coronagraphs remove the first $N$ electric-field modes. The lowest-order PC is a second-order coronagraph that removes the piston electric field mode, while a fourth-order coronagraph also removes the tip and tilt electric fields. The PC is defined as an orthogonal projection operation: 
\begin{equation}
E_{\mathrm{cor}} = E - \sum_{i=0}^{N-1} W_i <W_i, E>.
\end{equation}
Here, $E$ is the input electric field and $E_{\mathrm{cor}}$ is the output electric field. The PC removes modes $W_i$ from the input. The operator $<a,b>$ is the inner product between two functions $a$ and $b$. This operation is applied to each polarization state separately. While the PC does not exist in reality, there are several coronagraphs that closely follow the performance, such as the Phase Induced Amplitude Apodization Complex Mask Coronagraph for segmented apertures \citep{guyon2010high,belikov2018design} and the Vortex Coronagraph \citep{foo2005optical,mawet2005annular} for clear apertures. For other coronagraphs, the residuals depend on the particular design for each instrument. However, the dedicated high-contrast imaging instruments for the GMT \citep{males2022conceptual}, TMT \citep{fitzgerald2022planetary}, and ELT \citep{kasper2021pcs} are currently under development, and there are no coronagraph designs yet. The PC provides a fundamental limit on the performance of more realistic coronagraphs, which is why these have been applied in each case in this paper.

The coronagraphic residuals for the TMT are shown in Figure \ref{fig:coranagraphic_residuals_TMT}. They include a second-, fourth-, and sixth-order coronagraph in I band, H band, and L band. The results show that, after compensation of the common aberrations, a combination of diattenuation- and retardance-defocus is the dominant aberration. These are clearly visible in Figure \ref{jones-GSMT}. The defocus caused by the diattenuation is at a similar level to that of the phase aberration, which was found by simulations that considered only phase or amplitude aberrations in the Jones pupils. There is also some differential polarization beam shift (i.e., tip and/or tilt), which is visible in the second-order coronagraph residuals. The differential beam shift causes blurring of the stellar residuals. The change to a fourth-order coronagraph that also removes tip or tilt modes makes this very apparent; the nulls in the diffraction pattern are much sharper and deeper. The results show that polarization defocus is more important for these large and fast telescopes ($~$F/1) than polarization beam shifts, which are the dominant source of error in current ground-based telescopes (\cite{schmid2018sphere}, van Holstein in prep.). Higher-order modes barely play a role in the contrast budget. The sixth-order coronagraph residuals due to polarization aberrations are at a contrast level of $<10^{-8}$, which is well below the requirements of any ground-based direct imaging instrument. The downside of the sixth-order coronagraph is that its inner working angle  increases compared to the second- and fourth-order coronagraphs \citep{belikov2021theoretical}. Theoretically, it is possible to achieve an inner working angle of 1.5 $\lambda/D$ with an optimal  sixth-order coronagraph. This may be a sufficiently small enough inner working angle, depending on the exact requirements of the instrument. However, this is most likely not enough for GMagAO-X, which has an inner working angle goal of 1 $\lambda/D$. The coronagraphic residual images for the other two telescopes are similar in appearance and behavior and can be found in Appendix \ref{ap:corres}.

\begin{figure*}
\includegraphics[width=\textwidth]{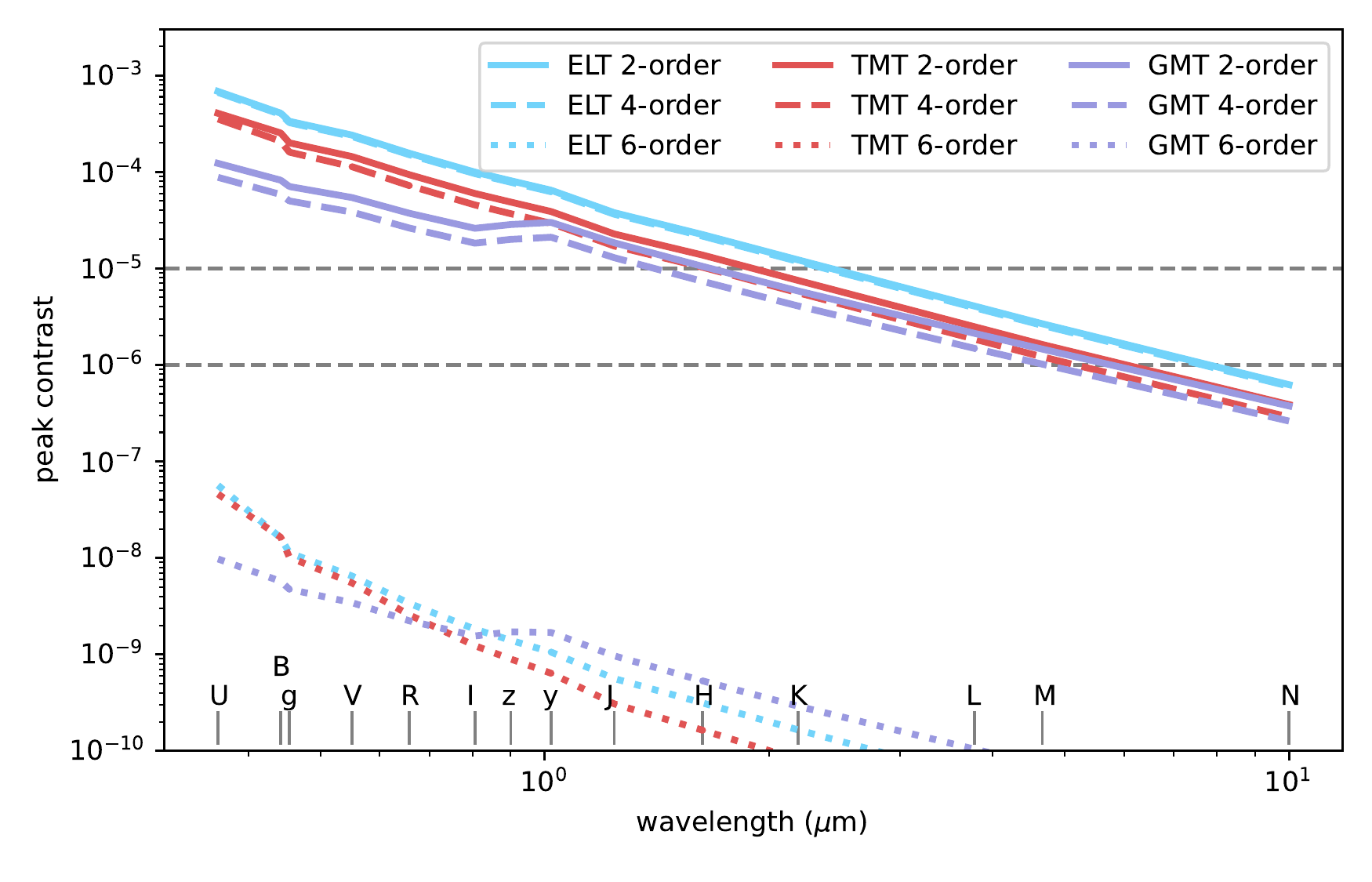}
\caption{Peak contrast as a function of wavelength and different coronagraphs for the ELT (blue), GMT (purple), and TMT (red). Each color represents a different telescope, and the line style changes for each coronagraph. The solid lines show the performance of a second-order coronagraph (theoretically ``perfect'' for an unaberrated system), the dashed lines that of a fourth-order coronagraph, and the dotted lines that of a sixth-order coronagraph. The center wavelength of each of the simulated photometric bands is shown at the bottom of the figure. The peak contrast shows an exponential decline with increasing wavelength. A peak contrast of $10^{-5}$ is reached between H and K band.}
\label{fig:peak_contrast_wavelength}
\end{figure*}

The polarization aberrations are wavelength dependent. The chromatic behavior is summarized by using the peak raw contrast in the residual images for each wavelength. The peak raw contrast as a function of wavelength is shown in Figure \ref{fig:peak_contrast_wavelength}. This figure shows the upper limit of the stellar leakage due to polarization aberrations. The results show that switching from a second-order coronagraph to a fourth-order barely affects the contrast. A sixth-order coronagraph would gain almost four orders of magnitude across all wavelengths. The peak contrast goes down as the wavelength increases, which was expected based on the chromatic behavior of the diattentuation and retardance. The expected raw contrast of the next generation of high-contrast imagers is about $10^{-5}$ at 1 $\lambda/D$ and is set by the performance of the AO systems \citep{kasper2021pcs,males2022conceptual}. Therefore, the polarization aberrations should be at most at the same contrast level, or they will be the dominant factor. The $10^{-5}$ level is reached for a wavelength around H band for the GMT and TMT. The ELT has worse performance since it has five mirrors instead of three and the fastest primary mirror. The ELT reaches the $10^{-5}$ level at wavelengths longer than  K band.

\begin{figure}
\includegraphics[width=\columnwidth]{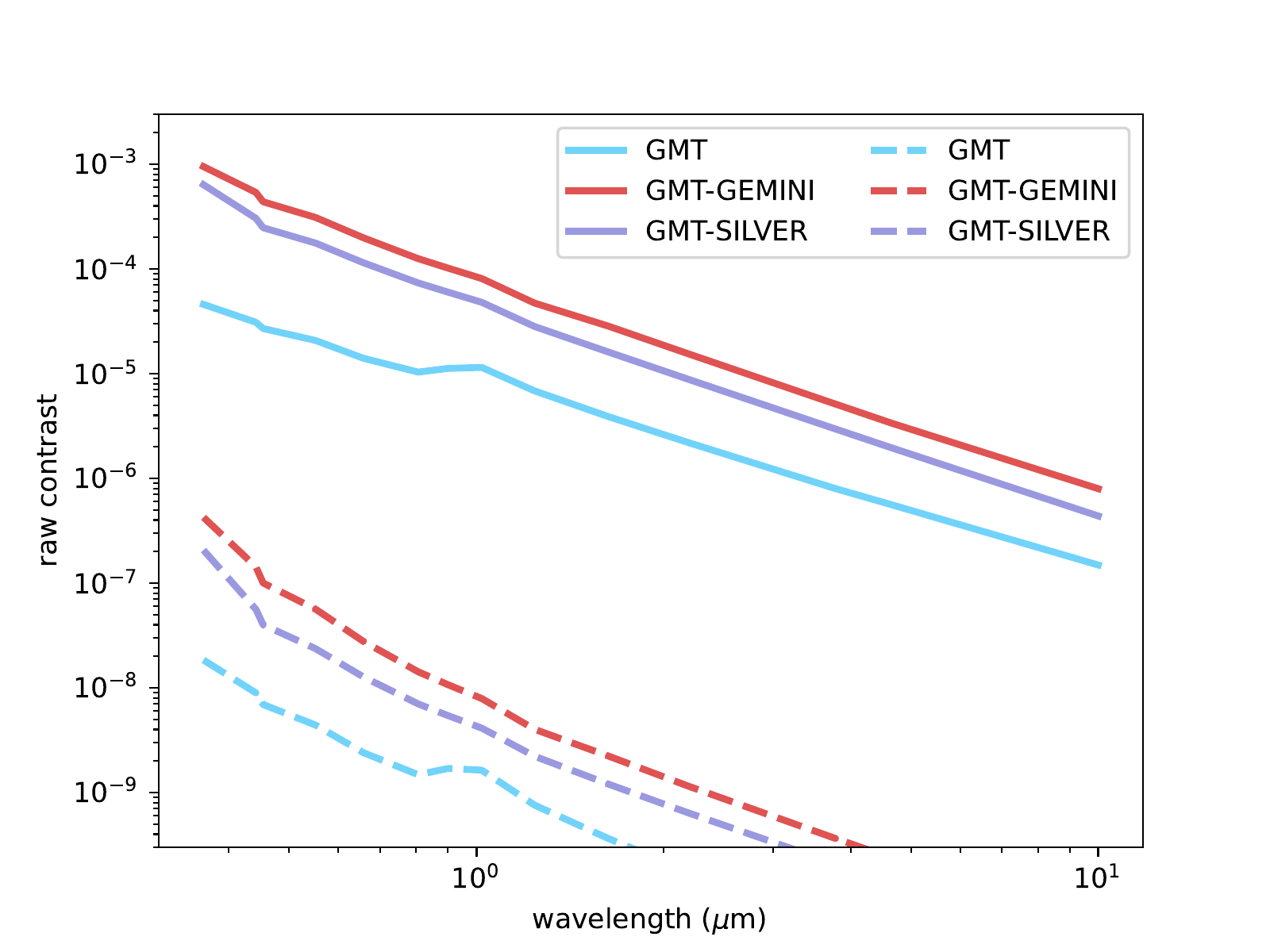}
\caption{Peak contrast as a function of wavelength for the GMT with different coatings. The different colors correspond to different mirror coatings. The solid lines show the contrast for a fourth-order coronagraph, and the dashed lines show the contrast for a sixth-order. The nominal aluminum coating for the GMT has the best performance. The difference in performance is even larger for the sixth-order coronagraph.}
\label{fig:gmt_coating_effects}
\end{figure}

The predicted raw contrast of the GMT is significantly better than that of the TMT, although both only have three mirrors. This could either originate from the geometric shape of the mirrors (incidence angles) or the coating specification (retardance and diattenuation). The simulations of the GMT were redone with two other coatings to disentangle the effects of geometry and coating. The first coating is identical to the TMT coating, and the second is a bare silver coating. The bare silver coating was investigated because the TMT coating is a multi-layer structure with silver as its base. The results can be seen in Figure \ref{fig:gmt_coating_effects}. The nominal aluminum coating of the GMT has the best performance. It outperforms the other coatings by an order of magnitude for both coronagraphs in the visible wavelength range. The aluminum coating is still better at longer wavelengths, but the gap between the coatings closes. These simulations demonstrate that the significant gain for the GMT over the other two telescopes is its coating. The performance gain is lost when the GMT uses the TMT coating. The contrast curves from the bare silver coating also show that the main cause of the polarization aberrations is the silver layer, which accounts for nearly 80\% of the residual contrast.

\section{Mitigation and calibration strategies}
\label{sec7}
There are several ways that the impact of polarization aberrations can be reduced or even completely removed.
\subsection{Coating of the mirrors}
The performance of the GMT is significantly better than that of the TMT, although both only have three mirrors. The main driver for the magnitude of the aberrations is the coating. The polarization aberrations could be reduced by optimizing the coating recipe not only for reflectivity, but also for retardance and diattenuation. One method for minimizing diattenuation and maximizing reflectivity was created for the Multiangle SpectroPolarimetric Imager (MSPI) \citep{Mahler2008}. The coatings of the MSPI mirrors were nominally protected silver coatings with two dielectric layers on top. The thickness of the two dielectric layers for each mirror was optimized against a merit function that weighted diattenuation and reflectivity equally. By doing so, the total diattenuation of the system was reduced to $<1 \%$ while maintaining a high reflectivity. A similar optimization could be conducted for the GSMT mirrors with additional consideration for retardance to mitigate the influence of polarization aberrations. As new coatings will be available in the future, this is a feasible approach for all the GSMTs as long as mirror coating facilities (the coating facility of the ELT and TMT has a modular design that is easy to upgrade) are designed to accommodate upgrades and delivery of different coatings \citep{schotsaert2020coating}.
\subsection{Compensation optics}
The straightforward approach for reducing polarization aberrations is the design of mirrors with an optimized curvature radius to minimize the incidence angle, avoiding fold mirrors in the optical configuration. Nevertheless, in all these GSMTs, fast primary mirrors have to be used to get to buildable telescope sizes, and the use of fold mirrors is inevitable because of instruments placed on the Nasmyth platform. Therefore, one of the mitigation techniques used to cancel the retardance tilt or beam shift from the fold mirror is using another  crossed-fold mirror with its $s-p$ planes rotated orthogonally to the M3 mirrors. However, this approach requires a trade-off study between the polarization aberration cancellation with the additional wavefront error and throughput loss.  
\citet{lam2015balancing} simulated the compensation of polarization aberrations using crossed-fold mirrors for a single field point. They obtain residual aberrations of linear variation of retardance and diattenuation, which is   easier to compensate with optimized orientations of the mirrors. A preliminary analysis of using crossed-fold mirrors to mitigate the instrumental polarization and crosstalk for TMT is explored by \citet{anche2018preliminary}, where IP reduces to 0.1\% from 4\%. Since our simulations show that the prominent polarization aberrations in these telescopes are retardance defocus and tilt, we could compensate for these using a spatially varying retarder optimized over wavelengths of interest. The design and analysis of the compensation optics and the calibration strategies will be explored in detail in the following papers in this series.
\subsection{Focal plane wavefront control}
The coronagraphic simulations were performed after a classical AO system. The AO systems optimize the wavefront for maximum Strehl by compensating the wavefront aberrations. For high-contrast observations, this is not necessarily optimal because of stellar speckles that leak through the coronagraph. Focal plane wavefront sensing and control can create dark holes in the PSF where the contrast is enhanced with respect to the planet. Typically, the electric field in the focal plane is estimated with some sensor (pair-wise probing or phase diversity), and then the DM is actuated to create an electric field that destructively interferes with all the light in a particular region. This approach to wavefront control is called electric field conjugation (EFC). EFC has been used on several testbeds to create very deep contrasts \citep{seo2019testbed, ruane2022broadband}.  It is now being implemented and tested for ground-based telescopes \citep{potier2022increasing, haffert2022advanced, ahn2022laboratory}.

Electric field conjugation has also been extended to multi-wavelength solutions and systems with phase and amplitude control by using multiple DMs. A similar approach with multiple DMs might make it possible to remove polarization aberrations. However, most EFC approaches have only been developed for scalar electric fields despite the fact that high-contrast testbeds are limited by polarization aberrations. The current testbeds try to reduce the influence of polarization aberrations by placing the coronagraphic arm of the instrument between polarizers to select only one polarization state. A single polarization state can be completely controlled. The downside to this approach is that half of the light is thrown away. This might not be an option for ground-based telescopes that still rely on advanced post-processing techniques that require high throughput. EFC can be extended to include control and sensing of both polarizations \citep{mendillo2021dual}. The main problem for EFC is disentangling the electric field of each polarization state during the estimation process. This is highly degenerate and makes the estimation process more difficult. The model-free approach of implicit EFC (iEFC) that only uses intensity-based measurement could be a solution to this \citep[][Haffert et al. in prep]{haffert2022advanced}. Future work will  show us whether the telescope-induced aberrations can be canceled to a deep enough level.

\subsection{Post-processing}
Any polarization aberration errors uncorrected by focal plane wavefront control will add a stellar photon noise floor, static or quasi-static speckles, to observations that cannot be completely removed.
Speckle subtraction techniques such as reference differential imaging (RDI) and angular differential imaging (ADI) allow imaging that approaches the photon noise floor.
The residual contrast due to the polarization aberrations discussed here should be largely static during observation, and thus well suited to removal by ADI \citep{marois2006angular}. In addition, RDI using PSF libraries images from other stars  \citep{soummer2011orbital,ruane2019reference} removes many of the observational constraints of ADI; however, through the life of an observatory, change to coatings may limit the effectiveness of RDI as will degenerate solutions to image plane wavefront control.
\section{Summary and conclusion}
\label{sec8}
The analysis of polarization aberrations is crucial for the next-generation GSMTs as they aim to reach an on-sky contrast of $10^{-5}$ to $10^{-4}$. Below, we summarize the significant results from our simulations of polarization aberrations.
\begin{enumerate}
    \item We estimated the polarization aberrations arising due to the telescope optics of next-generation GSMTs, including the segments of the primary mirror for all the astronomical bands.
    \item Our analysis indicates the presence of diattenuation defocus and tilt, and retardance defocuses and tilt as the prominent polarization aberrations, which cannot be corrected by the high-order adaptive optics instrument in these three telescopes.
    \item The peak raw contrast that can be achieved in these telescopes for the different order of the coronagraphs decreases with wavelength pointing to the correlation with the behavior of the mirror coating. The peak contrast in the blue region is $> 10^{-4}$, which is far below the requirements of high-contrast imaging instruments for these telescopes. Contrast better than $10^{-6}$ can be achieved only in the L, M, and N bands for the second-and fourth-order coronagraphs. 
    \item The raw contrast is estimated for the inner and outer working angle for the proposed high-contrast imaging instruments for each of the telescopes. For R and I bands the raw contrast at 1$\lambda/D$ for the second-and fourth-order coronagraphs is $>10^{-4}$ for ELT and TMT and  $>10^{-5}$ for GMT, which is an order less than the required contrast for the high-contrast imaging instruments.
    \item We also find that the performance of the aluminum coating is better than the Gemini-like coating, although the Gemini-like coating is optimized for higher reflectivity. This indicates the necessity to develop a coating optimization technique to incorporate the coating retardance and diattenuation in addition to the reflectivity.
    \item To overcome the beam shift caused by the retardance tilt of the fold mirror, it is crucial to design the compensation optics as a part of the high-contrast imaging instrument, especially considering the fact that only the telescope mirrors have been simulated here. Other elements inside the instruments themselves may generate polarization aberrations as well. The stellar residuals will become stronger if these are added to the telescope aberrations.
     \item As a part of this paper, we   developed  ZOS-API and Python-based polarization ray tracing routines  \citep{Ashcraft_poke_2022} that could be easily used to estimate the polarization aberrations of any optical system.
     \item The primary goal of direct imaging is to search for biomarkers with the GSMTs. One of the strongest bio-markers is the oxygen A band at 730 nm (between R and I bands), which requires a raw contrast of $10^{-5}$ to $10^{-4}$ \citep{snellen2015combining,lovis2022ristretto}. Unfortunately, the polarization aberrations arising from the telescope optics already introduce residuals at the required contrast levels. Additional aberrations from the instrument optics will only add to the current estimates. The polarization aberrations will need to be included in the coronagraph design of future HCI instruments.
\end{enumerate}

We plan to expand our simulations to include coating non-uniformity, coating aging, segment errors (including missing segments), realistic coronagraphs, and post-AO wavefront errors, including tip, tilt, and low-order aberrations and to  evaluate the polarization aberration structure and its statistical nature in the final PSF and post-processing residuals in the following paper.

\begin{acknowledgements}
SYH was supported by NASA through the NASA Hubble Fellowship grant \#HST-HF2-51436.001-A awarded by the Space Telescope Science Institute, which is operated by the Association of Universities for Research in Astronomy, Incorporated, under NASA contract NAS5-26555. This work was supported by a NASA Space Technology Graduate Research Opportunity. Portions of this work were supported by the Arizona Board of Regents Technology Research Initiative Fund (TRIF). Portions of this work were supported by the University of Arizona Space Institute’s seed grant for the GMT Pol instrument. The authors would like to thank Dr. Rebecca Bernstein (Giant Magellan Telescope project) and John Rogers (Thirty Meter Telescope) for the optical design files of the telescopes. The authors thank the anonymous referee who provided a constructive report that helped us improve this paper. The authors would like to dedicate this paper to Prof. James B. Breckinridge (1939-2022).
\end{acknowledgements}

\bibliographystyle{aa} 
\bibliography{ref}
\begin{appendix}
\section{Polarization aberration fitting of Jones pupils }
In this section we report on  a fit  to Jones pupils calculated for each telescope using the six analytical low-order polarization aberration terms: the diattenuation piston, tilt, and defocus ($d_0$, $d_1$, and $d_2$, respectively), and the retardance piston, tilt, and defocus ($\Delta_0$, $\Delta_1$, and $\Delta_2$, respectively). The terms represent (unitless) coefficients to the differential piston, tilt, and defocus terms that can be used to describe analytically a Jones pupil. A summary of these terms can be found in \cite{breckinridge2015polarization}, and a more detailed treatment can be found in \cite{chipman2018polarized} and references therein. 

For each telescope and each wavelength, we fit all six terms using a simple minimization of the root mean square of the diagonal terms of the Jones matrix. The off-diagonal terms were excluded because the discontinuities in the phase dominated the fitting and caused the fit to prefer nonphysical solutions. However, the on-diagonal terms contain contributions from all six terms, and thus could be used for the fit. Because of the large phase retardance introduced by M3 (i.e., retardance piston), we were unable to use the simplified version of the Jones pupil formalism presented in \cite{breckinridge2015polarization}, which assumed that each polarization aberration term was small in order to keep only first-order terms. Instead, we expanded the full cascade of Jones pupil terms $J_1; J_2; ...J_6$ keeping all orders of terms (see \citealt{chipman2018polarized} for the mathematical definition of each term). 
The results of our model fitting can be seen in Tables~\ref{tab:pol_aberration_tmt}, \ref{tab:pol_aberration_elt}, and \ref{tab:pol_aberration_gmt} for the TMT, ELT, and GMT, respectively. In all cases, the fits resulted in diagonal component amplitude residuals at less than the $10^{-3}$ level, and phase residuals of $10^{-3}$ radians or less, and often a magnitude or two lower. The off-diagonal components resulted in residuals at the $\sim10\%$ level as {compared to} the actual amplitude values, dominated by an artifact related to the phase discontinuities. The (small) diagonal residuals were dominated by higher-order terms that go beyond the standard low-order polarization aberration terms (up to defocus), but may be captured in future work if higher-order terms are included. Overall the fits represented the ray tracing results very well and the values in these tables can be used for a quick first-order estimate of the polarization aberrations. 

\begin{table*}
\begin{tabular}{ccccccc}
\hline \hline
Band & $d_0$ & $d_1$ & $d_2$ & $\Delta_0$ & $\Delta_1$ & $\Delta_2$ \\
\hline
U & 2.0e-02 & -7.0e-04 & -6.9e-03 & 1.0e+00 & -7.6e-02 & -2.1e-01 \\
B & 1.0e-02 & -7.0e-04 & -2.5e-03 & 7.2e-01 & -6.2e-02 & -1.4e-01 \\
V & 6.2e-03 & -5.0e-04 & -1.3e-03 & 5.4e-01 & -4.9e-02 & -1.0e-01 \\
g & 8.3e-03 & -6.2e-04 & -1.9e-03 & 6.4e-01 & -5.7e-02 & -1.2e-01 \\
R & 4.6e-03 & -4.1e-04 & -9.5e-04 & 4.3e-01 & -4.1e-02 & -8.2e-02 \\
I & 3.8e-03 & -3.5e-04 & -7.6e-04 & 3.4e-01 & -3.3e-02 & -6.5e-02 \\
z & 3.6e-03 & -3.4e-04 & -7.1e-04 & 3.1e-01 & -3.0e-02 & -5.8e-02 \\
y & 3.4e-03 & -3.2e-04 & -6.6e-04 & 2.8e-01 & -2.7e-02 & -5.2e-02 \\
J & 3.2e-03 & -3.1e-04 & -6.1e-04 & 2.1e-01 & -2.1e-02 & -4.0e-02 \\
H & 3.1e-03 & -3.0e-04 & -5.9e-04 & 1.6e-01 & -1.6e-02 & -3.1e-02 \\
K & 3.0e-03 & -3.0e-04 & -5.7e-04 & 1.2e-01 & -1.2e-02 & -2.3e-02 \\
L & 3.0e-03 & -3.0e-04 & -5.6e-04 & 6.9e-02 & -6.9e-03 & -1.3e-02 \\
M & 3.0e-03 & -2.9e-04 & -5.5e-04 & 5.6e-02 & -5.6e-03 & -1.1e-02 \\
N & 4.7e-03 & -4.7e-04 & -8.8e-04 & 2.7e-02 & -2.7e-03 & -5.1e-03 \\
\hline
\end{tabular}
\caption{Best-fit polarization aberration terms for TMT}
\label{tab:pol_aberration_tmt}
\end{table*}

\begin{table*}
\begin{tabular}{ccccccc}
\hline \hline
Band & $d_0$ & $d_1$ & $d_2$ & $\Delta_0$ & $\Delta_1$ & $\Delta_2$ \\
\hline
U & 2.0e-02 & -8.3e-04 & -1.1e-02 & 6.8e-01 & -4.5e-02 & -3.0e-01 \\
B & 7.6e-03 & -4.9e-04 & -3.6e-03 & 4.5e-01 & -3.3e-02 & -1.9e-01 \\
V & 4.4e-03 & -3.1e-04 & -1.9e-03 & 3.4e-01 & -2.6e-02 & -1.4e-01 \\
g & 6.0e-03 & -4.1e-04 & -2.8e-03 & 4.0e-01 & -3.0e-02 & -1.7e-01 \\
R & 3.2e-03 & -2.4e-04 & -1.4e-03 & 2.7e-01 & -2.1e-02 & -1.1e-01 \\
I & 2.6e-03 & -2.0e-04 & -1.1e-03 & 2.1e-01 & -1.7e-02 & -8.9e-02 \\
z & 2.5e-03 & -1.9e-04 & -1.0e-03 & 1.9e-01 & -1.5e-02 & -8.0e-02 \\
y & 2.3e-03 & -1.8e-04 & -9.8e-04 & 1.7e-01 & -1.4e-02 & -7.1e-02 \\
J & 2.2e-03 & -1.7e-04 & -9.0e-04 & 1.3e-01 & -1.1e-02 & -5.4e-02 \\
H & 2.1e-03 & -1.7e-04 & -8.7e-04 & 1.0e-01 & -8.2e-03 & -4.2e-02 \\
K & 2.0e-03 & -1.7e-04 & -8.5e-04 & 7.5e-02 & -6.1e-03 & -3.1e-02 \\
L & 2.0e-03 & -1.6e-04 & -8.3e-04 & 4.3e-02 & -3.5e-03 & -1.8e-02 \\
M & 2.0e-03 & -1.6e-04 & -8.3e-04 & 3.5e-02 & -2.8e-03 & -1.4e-02 \\
N & 2.8e-03 & -2.2e-04 & -1.1e-03 & 1.7e-02 & -1.4e-03 & -6.9e-03 \\
\hline
\end{tabular}
\caption{Best-fit polarization aberration terms for ELT}
\label{tab:pol_aberration_elt}
\end{table*}
\begin{table*}
\begin{tabular}{ccccccc}
\hline \hline
Band & $d_0$ & $d_1$ & $d_2$ & $\Delta_0$ & $\Delta_1$ & $\Delta_2$ \\
\hline
U & -2.7e-02 & -4.7e-03 & 9.1e-03 & -3.2e-01 & -5.6e-02 & 1.0e-01 \\
B & -2.8e-02 & -5.0e-03 & 9.3e-03 & -2.6e-01 & -4.7e-02 & 8.2e-02 \\
V & -3.1e-02 & -5.5e-03 & 1.0e-02 & -2.1e-01 & -3.8e-02 & 6.7e-02 \\
g & -2.9e-02 & -5.2e-03 & 9.5e-03 & -2.4e-01 & -4.4e-02 & 7.7e-02 \\
R & -3.5e-02 & -6.3e-03 & 1.1e-02 & -1.8e-01 & -3.2e-02 & 5.6e-02 \\
I & -4.9e-02 & -8.8e-03 & 1.6e-02 & -1.5e-01 & -2.8e-02 & 4.8e-02 \\
z & -4.7e-02 & -8.5e-03 & 1.5e-02 & -1.6e-01 & -2.9e-02 & 5.0e-02 \\
y & -2.6e-02 & -4.7e-03 & 8.3e-03 & -1.5e-01 & -2.8e-02 & 4.9e-02 \\
J & -1.2e-02 & -2.2e-03 & 3.9e-03 & -1.1e-01 & -2.1e-02 & 3.6e-02 \\
H & -8.7e-03 & -1.6e-03 & 2.8e-03 & -8.5e-02 & -1.6e-02 & 2.7e-02 \\
K & -7.7e-03 & -1.4e-03 & 2.5e-03 & -6.3e-02 & -1.2e-02 & 2.0e-02 \\
L & -6.4e-03 & -1.2e-03 & 2.0e-03 & -3.7e-02 & -6.9e-03 & 1.2e-02 \\
M & -5.8e-03 & -1.1e-03 & 1.8e-03 & -3.1e-02 & -5.7e-03 & 9.8e-03 \\
N & -4.4e-03 & -8.1e-04 & 1.4e-03 & -1.5e-02 & -2.8e-03 & 4.8e-03 \\
\hline
\end{tabular}
\caption{Best-fit polarization aberration terms for GMT}
\label{tab:pol_aberration_gmt}
\end{table*}

\section{Coronagraphic residuals}
\begin{figure*}
\includegraphics[width=\textwidth]{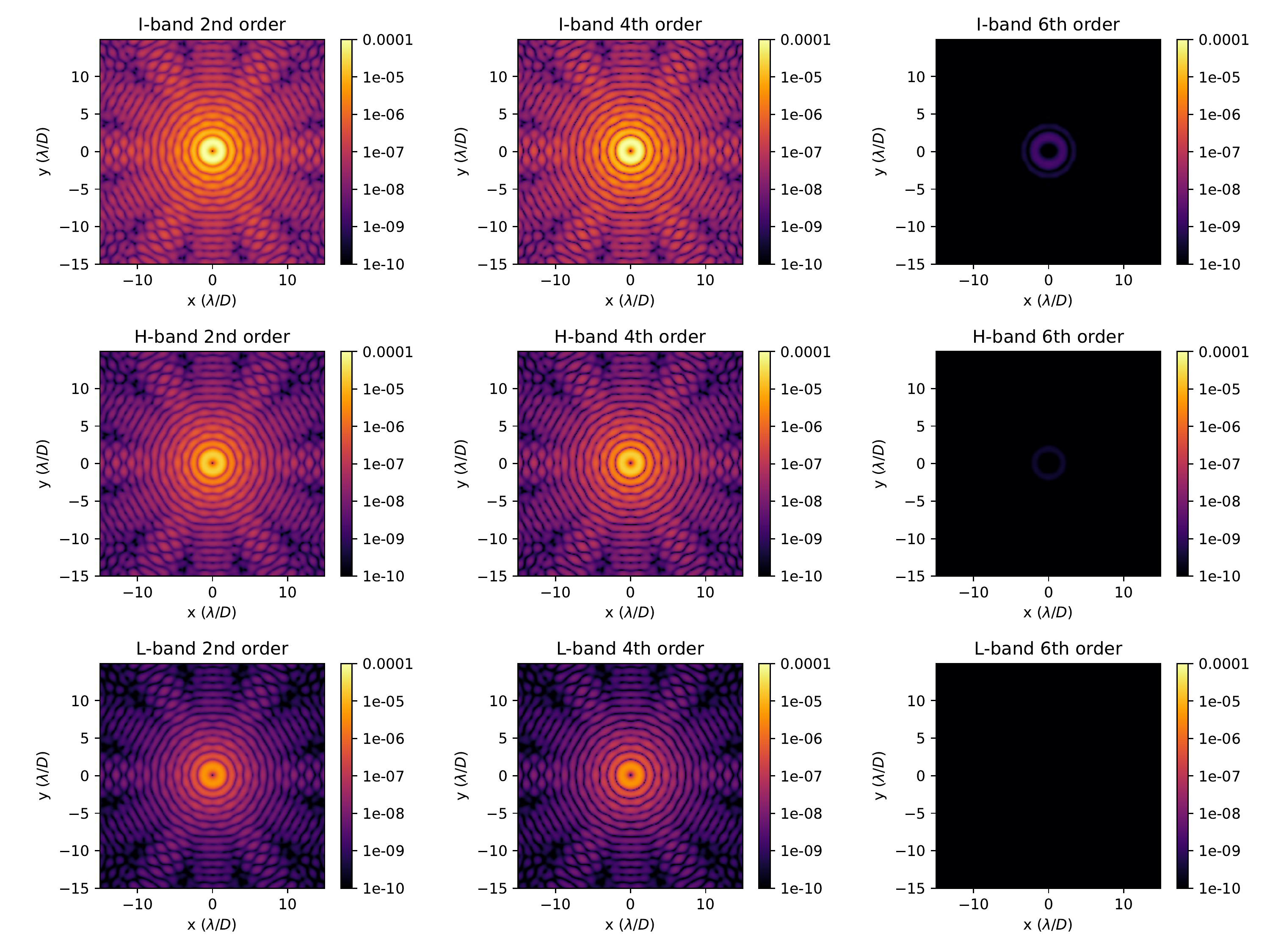}
\caption{Stellar residuals for different wavelengths and coronagraphs for the ELT. The residuals are shown for I band (top), H band (middle), and L band (bottom). The coronagraphs are second- (left), fourth- (middle), and sixth-order (right) coronagraphs. These images show that the stellar residuals are mainly defocus.}
\label{fig:coranagraphic_residuals_ELT}
\end{figure*}

\begin{figure*}
\includegraphics[width=\textwidth]{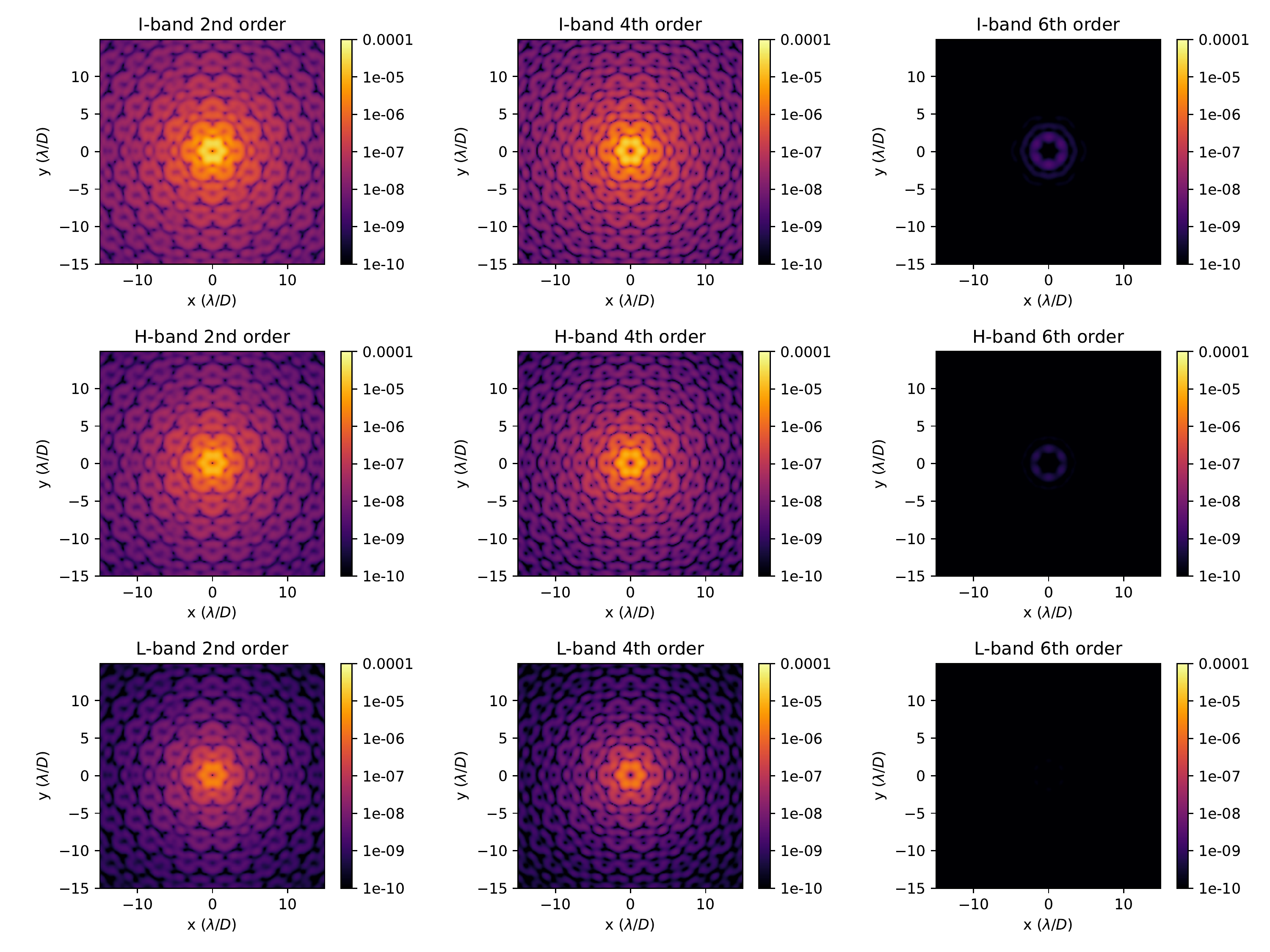}
\caption{Stellar residuals for different wavelengths and coronagraphs for the GMT. The residuals are shown for I band (top), H band (middle), and L band (bottom). The coronagraphs are second- (left), fourth- (middle), and sixth-order (right) coronagraphs. These images show that the stellar residuals are mainly defocus.}
\label{fig:coranagraphic_residuals_GMT}
\end{figure*}

The coronagraphic residuals due to polarization aberrations for the ELT and GMT are shown in Figures \ref{fig:coranagraphic_residuals_ELT} and \ref{fig:coranagraphic_residuals_GMT}.
\label{ap:corres}

\section{Refractive indices of coatings used in our simulations}
\label{sec:Appendix}
Table \ref{tab:refractive_indices} provides the list of refractive indices for different coatings used in our simulations for all the astronomical bands. 
\begin{table*}
\begin{center}
\begin{tabular}{cccccccc}
\hline
     &            & \multicolumn{2}{c}{Ag (\cite{yang2015})} & \multicolumn{2}{c}{Al (\cite{rakic1995algorithm})} & \multicolumn{2}{c}{Si3n4 (\cite{luke2015broadband,kischkat2012mid})} \\
Band & Wavelength ($\mu$m) & n           & k        & n          & k         & n        & k            \\
\hline
U    & 0.364      & 0.070917    & 1.6019   & 0.39732    & 4.3875    & 2.1229   & 0            \\
B    & 0.442      & 0.053192    & 2.5332   & 0.60555    & 5.3577    & 2.0815   & 0            \\
g    & 0.475      & 0.051893    & 2.8414   & 0.72122    & 5.7556    & 2.0703   & 0            \\
V    & 0.54       & 0.052824    & 3.4018   & 0.97274    & 6.5119    & 2.0543   & 0            \\
R    & 0.647      & 0.060334    & 4.2604   & 1.5389     & 7.6818    & 2.0377   & 0            \\
I    & 0.789      & 0.077492    & 5.3462   & 2.7233     & 8.4171    & 2.0249   & 0            \\
z    & 0.866      & 0.089423    & 5.9221   & 2.4499     & 8.1439    & 2.0201   & 0            \\
y    & 0.962      & 0.1065      & 6.6335   & 1.53       & 8.9597    & 2.0154   & 0            \\
J    & 1.24       & 0.1686      & 8.664    & 1.3158     & 12.246    & 2.0053   & 0            \\
H    & 1.63       & 0.2848      & 11.48    & 1.7022     & 16.502    & 2.4618   & 0.00004      \\
K    & 2.19       & 0.50862     & 15.487   & 2.732      & 22.185    & 2.4522   & 0.0001       \\
L    & 3.78       & 1.4983      & 26.757   & 6.2634     & 36.771    & 2.4036   & 0.00082623   \\
M    & 4.66       & 2.2661      & 32.936   & 8.3289     & 44.335    & 2.3558   & 0.0025856    \\
N    & 10         & 10.006      & 69.039   & 25.006     & 85.965    & 1.627    & 1.1541      \\
\hline
\end{tabular}
\end{center}
\caption{Refractive indices used in our calculations}
\label{tab:refractive_indices}
\end{table*}
\end{appendix}
\end{document}